\documentclass[
    aps,
    pra,
    onecolumn,
    superscriptaddress,
    nofootinbib,
    amsmath,
    amssymb
]{revtex4-2}

\usepackage[T1]{fontenc}
\usepackage[utf8]{inputenc}
\usepackage{quantikz}

\usepackage{xcolor}
\usepackage{indentfirst}
\usepackage{svg}
\usepackage{tikz}
\usepackage{graphicx}
\usepackage[percent]{overpic}
\usepackage{braket}
\usepackage{hyperref}
\usepackage{setspace}
\hypersetup{
    colorlinks=true,
    linkcolor=blue,
    filecolor=magenta,
    urlcolor=blue,
    citecolor=blue,
    pdfborder={0 0 0},
    pdfborderstyle={},
}

\begin{document}


\title{Noise structuring in fixed-depth Trotter simulation: stationary channels and observable-level depolarization}

\author{G. L. Stavisskii}
\affiliation{Dukhov Research Institute of Automatics (VNIIA), Moscow, 127055, Russia}
\affiliation{Moscow Institute of Physics and Technology, Dolgoprudny, 141700, Russia}

\author{W. V. Pogosov}
\affiliation{Dukhov Research Institute of Automatics (VNIIA), Moscow, 127055, Russia}
\affiliation{Institute for Theoretical and Applied Electrodynamics, \\ Russian Academy of Sciences, Moscow, 125412, Russia}

\author{L. E. Fedichkin}
\affiliation{Dukhov Research Institute of Automatics (VNIIA), Moscow, 127055, Russia}
\affiliation{Moscow Institute of Physics and Technology, Dolgoprudny, 141700, Russia}

\author{A. V. Lebedev}
\affiliation{Dukhov Research Institute of Automatics (VNIIA), Moscow, 127055, Russia}
\affiliation{Moscow Institute of Physics and Technology, Dolgoprudny, 141700, Russia}

\begin{abstract}
In noisy quantum simulations, distinguishing intrinsic coherent evolution from distortions caused by hardware noise is a central challenge. In conventional time-dependent Trotter simulations, noise accumulation varies with the endpoint time, obscuring dynamical observables and requiring substantial error mitigation overhead for long simulations. To address those issues, we analyze fixed-depth Trotter simulation for structuring hardware noise in digital many-body dynamics. Fixing the layer count—chosen via the largest endpoint time—throughout the time scan makes the total noise dose approximately independent of the endpoint time. For local stochastic faults, we show that, once propagated faults lose memory of their insertion layer, the noisy circuit factorizes into the ideal evolution followed by a stationary finite-depth binomial channel. The memory time of a single fault is related to a Loschmidt echo. Noise produces observable-level depolarization resulting in an almost time-independent affine contrast correction for  macroscopic observables at low-to-moderate noise, crucial for error mitigation. At short times, the protocol produces a digital Zeno-like transient. Finally, our results reveal limitations of naive zero-noise extrapolation using oversimplified functions.
\end{abstract}

\maketitle


\section{Introduction}

Digital quantum simulation aims to reproduce real-time many-body dynamics on gate-based processors. Such simulations probe transport, correlation spreading, relaxation after quenches, scrambling, and dynamical phase transitions \cite{feynman1982, excitation_transport_ion_comp, google_quantum_OTOC, dynamic_phase_transitions_exp, Polkovnkiov_dynamical_phase_transitions, Gibbs_closed_QS_exp, quantum_correlations_2014}, including regimes that are generally difficult to simulate classically \cite{Preskill_2018, review_sim, lloyd_universal_qsim}. Recent superconducting, trapped-ion, and Rydberg experiments have made digital and hybrid simulations increasingly concrete \cite{Ion_ising_2026, 2025_ion_fermionic, 2025_supercond_pairing_corr_ion, qdps_supercond, Keenan_2023, Mueller_2025, rydberg_qcd_string, scattering_IBM, Trotter_VNIAA_exp, Hybrid}. The standard implementation uses Trotter--Suzuki product formulas \cite{review_sim, babbush2017}: increasing the number of layers improves the mathematical approximation but also increases the number of noisy gates \cite{Areg}.

In general, incoherent noise can significantly distort the dynamics of observables of interest. In the NISQ regime, where fully fault-tolerant quantum error correction is not yet available, various error mitigation techniques are used to recover information about the underlying ideal dynamics, such as zero-noise extrapolation (ZNE) \cite{ZNE_intro} and probabilistic error cancellation (PEC) \cite{PEC2017}. Nevertheless, their application to long-time dynamical simulations remains challenging for current protocols. ZNE requires obtaining data at several noise levels for each time point of interest. In conventional Trotter simulations, different endpoint times are typically represented by circuits with different depths, so the corresponding noise scaling and extrapolation procedures must be repeated throughout the dynamical trajectory, with sampling overhead growing exponentially with algorithm depth. PEC relies on accurate characterization of the implemented noise processes and suffers from an exponential sampling overhead that scales with the number of noisy operations in the circuit \cite{Endo2018}, which becomes increasingly challenging for large-scale dynamical scans involving deep circuits. Thus, both widely used mitigation techniques lead to the bottleneck of exponentially growing overhead. It is desirable to develop protocols without excessive mitigation overhead across the entire dynamical trajectory that will enable the study of long-time dynamics by alleviating the gate-complexity bottleneck.

A useful error mitigation picture is that a noisy circuit may be viewed as the ideal circuit dressed by an effective global noise channel \cite{Global_noise_rev, Qin_2023_Global_noise, Devitt_2013}. Sufficiently random circuits can approach a global white-noise form \cite{dalzell2024}, whereas shallow and structured circuits need not do so automatically \cite{Foldager_2024_white_noise}. Noisy Trotterized dynamics can also be described through algorithm-dependent static effective Lindbladians \cite{Fratus_2025_static_lindblad}. Here, we ask a complementary question motivated by the problems of error mitigation: can a structured Trotter circuit, scanned at fixed depth, generate an approximately endpoint-independent noise channel for the measured dynamics?

The answer depends on the time-scan protocol. In a fixed-step scan, the Trotter step is kept constant and the circuit depth grows with endpoint time. In a fixed-error scan, the step size is adjusted at each endpoint to keep the product-formula error approximately fixed. The fixed-depth protocol studied here has a different purpose: the number of layers is chosen once from the Trotter-accuracy requirement at the largest endpoint time in a target window and is then held fixed throughout the scan. This is not pointwise noise-optimal at short times, but it makes the total number of noisy locations approximately endpoint independent.

Our main result is a stationary-channel description of this fixed-depth regime, which provides a controlled separation between the noise contribution and the unitary dynamics contribution. A local stochastic fault inserted at an unknown Trotter layer is propagated by the target dynamics. Once its averaged effect has lost memory of the insertion position, it defines a stationary single-fault channel. For independent faulty layers, summing over their number gives a finite-depth binomial stationary global channel, which reduces to the Poissonian exponential form in an additional dilute-layer limit. The memory loss used here is operational: convergence of an averaged channel is naturally related to a Loschmidt echo of a local perturbation.

A particularly useful consequence appears at the level of observables. The stationary global channel is generally not completely depolarizing on the full operator algebra. Nevertheless, for selected macroscopic observables at moderate noise levels, its action on the relevant data set can be well approximated by an endpoint-independent affine transformation, namely a contrast factor plus offset. Similar behavior has previously been observed both in experiments on quantum hardware \cite{Trotter_VNIAA_exp} and in classical simulations of Trotterized quantum circuits with Pauli noise \cite{babukhin2022effect}. This regime is very attractive for quantum error mitigation, since it allows for a simple reconstruction of the error-free dynamics; this partly addresses the aforementioned problems of classical mitigation strategies. In particular, in order to achieve this simplification, a fixed-depth protocol can be used together with the protocols discussed above and classical estimates of the affine-transformation
parameters.  
However, at shorter endpoint times, the same fixed-depth protocol exhibits a digital Zeno-like transient: the coherent rotation angle per layer vanishes while the number of noise opportunities remains finite, so stochastic interruptions can dominate the coherent motion they interrupt \cite{misra1977zeno, Zeno_Fedichkin, itano1990quantumzeno, facchi2008quantumzeno, Fischer_2001, zeno_subspace_exp}. This result shows that zero-noise extrapolation approaches based on oversimplified functions can produce inadequate results when the noise level drives the system across the crossover into the Zeno-like regime.  

The paper is organized as follows. Section~\ref{sec:protocol} introduces the fixed-depth protocol and the noise-balance argument. Section~\ref{sec:single_fault} derives the single-fault channel and its Loschmidt-echo memory time. Section~\ref{sec:poisson} derives the binomial global channel and its Poisson limit. Section~\ref{sec:rescaling} discusses effective affine rescaling and the influence of the dynamics on ZNE together with the gate overhead required by the fixed-depth protocol. Section~\ref{sec:conclusion} summarizes the results.

\section{Fixed-depth Trotter protocol as a noise-structuring scheme}
\label{sec:protocol}

This section defines the fixed-depth protocol and the stochastic error model. We emphasize the practical role of the protocol from the outset: it is a noise-structuring choice. The goal is not to minimize the number of gates at each endpoint time, but to make the leading hardware-noise channel as uniform as possible over a time scan. This uniformity allows microscopic faults, after they have lost memory of their insertion positions, to be represented by a nearly stationary channel. The resulting stationary rescaling, based on an observable-level depolarization approximation, is then controlled over a range of noise strengths.

\subsection{Trotterized dynamics}

We consider the digital simulation of a many-body Hamiltonian with local interactions. For concreteness, and also for the numerical examples below, one may keep in mind the one-dimensional transverse-field Ising model with periodic boundary conditions. The main arguments do not rely on this particular Hamiltonian; they use locality and the product-formula structure of the digital simulation. The Hamiltonian reads as
\begin{equation}
    H_{TFIM}=H_A+H_B,
    \label{eq:H_AB}
\end{equation}
where
\begin{equation}
    H_A=\sum_{j=1}^{N_{\rm q}} h_j X_j,
    \qquad
    H_B=\sum_{j=1}^{N_{\rm q}} J_j Z_jZ_{j+1}.
    \label{eq:ising_AB}
\end{equation}
Here $N_{\rm q}$ is the number of qubits, $h_j$ is the local transverse field, and $J_j$ is the nearest-neighbor coupling. We use dimensionless units, with all energies measured in units of $\bar{J} = \sum_{j = 1}^{N_{q}}J_{j} / N_{q}$ and time measured in units of $1 / \bar{J}$.

The exact evolution to an endpoint time $T$ is controlled by the evolution operator $U(T)=e^{-iHT}$, where $H$ and $T$ are the dimensionless Hamiltonian and time, respectively. In a first-order Trotter approximation, it is replaced by $N$ identical layers,
\begin{equation}
    U(T)\simeq U_{\Delta}^{N},
    \qquad
    \Delta=\frac{T}{N},
    \label{eq:fixed_N_delta}
\end{equation}
where
\begin{equation}
    U_{\Delta}=e^{-iH_{A}\Delta}e^{-iH_{B}\Delta}.
    \label{eq:first_order_layer_AB}
\end{equation}

At finite $N$, the ideal channel in the noise-factorization formulas below should be understood as the noiseless Trotter channel generated by the chosen product formula. In the continuous-step limit this channel approaches the exact Hamiltonian evolution. Mathematical Trotter error due to the discretization and stochastic hardware noise are therefore kept conceptually separate throughout the discussion.

In the Ising example, the layer is implemented through one-qubit rotations generated by $X_j$ and two-qubit rotations generated by $Z_jZ_{j+1}$. The latter can be compiled using CNOT gates or implemented using native two-qubit rotations such as $R_{ZZ}$ or $R_{ZX}$. In the main text, we discuss only the first-order Trotterization scheme with CNOT transpilation; the effects of higher-order Trotterizations and alternative transpilations are discussed in the appendices.

\subsection{Complementarity of protocols}

Suppose that the dynamics is required on an interval $0\le T\le T_{\max}$. A standard fixed-step scan keeps the Trotter step fixed and changes the number of layers with time. This is efficient at short times and is a natural way to generate a uniformly sampled trajectory. A fixed-error scan goes further and adjusts the step size with the endpoint time in order to keep the coherent product-formula error approximately constant. Both protocols are useful, and both are natural benchmarks for digital simulation.

The fixed-depth protocol is motivated by a different practical objective. The number of layers $N$ is chosen from the accuracy requirement at the largest endpoint time $T_{\max}$ and then kept fixed for all smaller endpoint times. The mathematical Trotter error therefore decreases at shorter times, while the number of noisy layers remains the same. This deliberately sacrifices short-time noise optimality in order to make the total hardware-noise dose approximately independent of endpoint time. Here, "fixed depth" means that the same compiled gate skeleton is executed throughout the scan. 

This tradeoff can be advantageous when the goal is to reconstruct an extended time trace. In a conventional scan, the noise strength changes with the number of layers and therefore with time. In the fixed-depth scan, by contrast, the leading noise channel can become almost the same for all endpoint times after the short-time crossover. The resulting correction is then low-dimensional: for a given observable one needs only a contrast factor plus a possible offset, rather than a separate noise model for every time point.

Thus, the fixed-depth protocol is not a replacement for fixed-step or fixed-error simulations; it should be regarded as a complementary mode of operation. Fixed-step and fixed-error data are well suited for checking Trotter convergence, resolving early-time behavior, and reducing unnecessary depth at selected endpoints. Fixed-depth data are useful for identifying a long-time window in which the hardware noise is nearly stationary and can be calibrated globally. Agreement between corrected fixed-depth data and data obtained from other protocols provides a nontrivial consistency check on both the Trotter approximation and the noise-mitigation procedure.


\subsection{Local stochastic error model}

We describe circuit imperfections using a local stochastic
Pauli-error model. After each elementary one-qubit gate,
a depolarizing channel acts on the corresponding qubit,
\begin{equation}
    D_p^{(1)}(\hat{\rho})
    =
    (1-p)\hat{\rho}+\frac{p}{3}
    \left(X\hat{\rho}X+Y\hat{\rho}Y+Z\hat{\rho}Z\right),
    \label{eq:one_qubit_depol}
\end{equation}
where $p$ is the effective error probability at one
elementary noise location. For a two-qubit gate, we assume
independent depolarizing errors on the participating qubits,
\begin{equation}
    D_p^{(2)}=D_p^{(1)}\otimes D_p^{(1)}.
    \label{eq:two_qubit_depol}
\end{equation}
This differs from uniform depolarization over all
15 nonidentity two-qubit Pauli operators.

The stochastic Pauli description is motivated partly by
randomized compiling (RC), which can tailor a broad class
of physical noise channels, including coherent errors,
into effective stochastic Pauli channels under suitable
assumptions~\cite{WallmanEmerson2016}. Its effectiveness
has also been demonstrated experimentally on superconducting
processors~\cite{HashimEtAl2021}. However, RC does not
generally produce isotropic depolarizing noise or eliminate
spatial and temporal correlations.

Importantly, effective depolarizing noise does not imply
neglecting relaxation and dephasing associated with finite
$T_1$ and $T_2$. These processes contribute to the
effective error probability. Pauli twirling yields effective
stochastic Pauli errors, generally with unequal
probabilities, while the depolarizing approximation
replaces them by a single parameter $p$. This retains
their effective stochastic contribution without reproducing
all microscopic features, such as the nonunital component
of amplitude damping. The relation between $p$ and
physical coherence times also depends on gate durations
and other error sources.

Furthermore, recent experimental evidence suggests
that local depolarizing noise can effectively model
more complex physical error mechanisms, including
relaxation, dephasing, and gate imperfections,
while yielding similar results for relevant
many-body observables~\cite{google_quantum_OTOC}.

Thus, our analytical framework is not fundamentally restricted
to depolarizing noise. The derivations rely on decomposing
stochastic errors into Pauli fault configurations and
can accommodate arbitrary stochastic Pauli channels
by modifying the statistical weights of fault patterns.
This includes biased, gate-dependent, and correlated
Pauli errors, although correlations may require revising
the fault statistics below.

Let $n_{\rm loc}$ be the number of elementary depolarizing
locations in one Trotter layer, counting both local
channels following each two-qubit gate. Assuming
independent faults with uniform probability $p$, the
probability of at least one fault in a layer is
\begin{equation}
    q=1-(1-p)^{n_{\rm loc}},
    \label{eq:layer_error_probability_q}
\end{equation}
and the mean number of faulty layers in a depth-$N$
circuit is
\begin{equation}
    \mu=Nq.
    \label{eq:mu_def_new}
\end{equation}
For $pn_{\rm loc}\ll1$, one has $q\simeq pn_{\rm loc}$,
but we retain the exact expression when evaluating
layer-level fault statistics. For independent and
identically distributed errors, the number $M$ of
faulty layers follows the binomial distribution
\begin{equation}
    \Pr(M)=\binom{N}{M}q^M(1-q)^{N-M}.
    \label{eq:binomial_fault_distribution_intro}
\end{equation}
The Poisson distribution with mean $\mu$ arises only
in the additional limit $q\to0$, $Nq=\mu$ fixed.
Nonuniform but independent layer probabilities instead
give a Poisson-binomial distribution, while correlated
errors require more general joint fault statistics.

It is useful to distinguish two smallness assumptions. The first is microscopic locality of the elementary fault patterns, controlled by the local error probabilities $p$. The second is diluteness on the scale of whole Trotter layers, $q\ll1$. The exact binomial formulas do not require replacing the binomial distribution by a Poisson distribution. However, some continuum ordered-sum estimates and the leading approximation in which intra-layer coherent corrections are neglected become most transparent in the dilute-layer regime. In a thermodynamic limit at fixed local $p$, $q=1-(1-p)^{n_{\rm loc}}$ may approach unity as $n_{\rm loc}$ grows. Then the layer-level dilute description should be replaced by a local space-time fault-density formulation. The present work focuses on the finite-size, dilute-layer regime relevant for the circuits analyzed numerically, while retaining the binomial channel as the primary finite-$N$ expression.

\subsection{Zeno-like noise balance and short-time transient}

The coherent rotation angle accumulated in one Trotter layer is proportional to the step size. If $\Omega$ denotes a characteristic frequency of the target dynamics relevant for a given observable, then the coherent angle per layer has the scale
\begin{equation}
    \theta(T)\sim \Omega\Delta=\frac{\Omega T}{N}.
    \label{eq:coherent_angle_layer_new}
\end{equation}
The stochastic error dose per layer is approximately $q$. Therefore the noise-to-coherent-rotation ratio in one layer is
\begin{equation}
    \frac{q}{\theta(T)}
    \sim
    \frac{Nq}{\Omega T}
    =
    \frac{\mu}{\Omega T}.
    \label{eq:noise_to_rotation_ratio_new}
\end{equation}
For short endpoint times, this ratio is large: the circuit applies many noisy layers, each producing only a very small coherent rotation, and stochastic errors can repeatedly destroy or randomize the coherences that the Hamiltonian attempts to build. We use the term Zeno-like in this work in a phenomenological and operational sense. The control parameter is the imbalance $q/\theta(T)$: in a fixed-depth scan, the number of noise opportunities remains fixed as $T\to0$, whereas the coherent angle generated by each layer vanishes as $T/N$. This does not mean that the circuit implements projective measurements. Rather, the same physical competition found in dephasing- or measurement-induced Zeno suppression is present: incoherent interruptions dominate the coherent motion \cite{misra1977zeno, Zeno_Fedichkin, itano1990quantumzeno, facchi2008quantumzeno, Fischer_2001, zeno_subspace_exp}. Operationally, the transient is the part of the fixed-depth trace in which the stationary-channel correction has not yet formed. Applying a single affine stationary rescaling in this regime can even produce values outside the physical range of the observable. Here such behavior is treated as a useful diagnostic of the breakdown of the stationary-rescaling window, not as a physical magnetization value.

As $T$ increases, the coherent angle per layer grows while the error probability per layer stays approximately fixed. The system leaves the Zeno-like transient when the coherent motion becomes strong enough on the scale of a single layer. A simple estimate gives
\begin{equation}
    \Omega t_{\rm cross}/N\sim q,
    \qquad
    t_{\rm cross}\sim \frac{\mu}{\Omega}.
    \label{eq:t_cross_simple}
\end{equation}
In a many-body system, $\Omega^{-1}$ should be replaced by the memory time of a local fault. In the derivation below this time is denoted by $T^\ast$ and is estimated from a Loschmidt-echo-type quantity. The corresponding sufficient time scale for the formation of a stationary global noise channel is
\begin{equation}
    T_{\rm stat}\sim \mu T^\ast=NqT^\ast.
    \label{eq:Tstat_physical}
\end{equation}
This scaling is the basic scale estimate of the noise-balance picture.

\subsection{Endpoint-Lindblad picture}

In this subsection, we reconsider the same fixed-depth Trotterized evolution in a continuum approximation based on an effective Lindblad picture. The purpose is not to introduce an independent microscopic noise model, but to show, on a qualitative level, how the same noise-balance argument appears when the discrete circuit is viewed as an auxiliary open-system evolution. This discussion is closely related in spirit to effective-Lindbladian descriptions of noisy Trotterized dynamics, where circuit-level noise can be reinterpreted as static, algorithm-dependent dissipative terms acting in addition to the target Hamiltonian evolution \cite{Fratus_2025_static_lindblad}. The distinction is that the equation below is not a single physical master equation shared by all endpoint times. It is an auxiliary description of a family of fixed-depth circuits, and its effective rate changes with the endpoint time.

For a fixed endpoint time $T$ and a fixed number of Trotter layers $N$, the noisy circuit can be approximated by an auxiliary Lindblad evolution on an internal variable $s\in[0,T]$:
\begin{equation}
    \frac{d\hat{\rho}^{[T]}(s)}{ds}
    =
    \left[\mathcal L_H+\bar\gamma(T)\mathcal D\right]\hat{\rho}^{[T]}(s),
    \qquad 0\le s\le T,
    \label{eq:endpoint_lindblad_new}
\end{equation}
where $\mathcal L_H\hat{\rho}=-i[H,\hat{\rho}]$ and $\mathcal D$ is an effective dissipator determined by the microscopic fault model. Since a noise dose $q$ is accumulated over a simulated time step $\Delta = T/N$, dimensional reasoning gives
\begin{equation}
    \bar\gamma(T)\sim \frac{q}{\Delta}=\frac{Nq}{T}=\frac{\mu}{T}.
    \label{eq:endpoint_gamma_new}
\end{equation}
It is essential that $\bar\gamma(T)$ is not a time-local rate of one continuous trajectory valid for all endpoint times. It is constant along the auxiliary interval $0\le s\le T$ for a fixed endpoint $T$, but it changes when a different endpoint time is simulated.

Equation~\eqref{eq:endpoint_gamma_new} explains why an approximately time-independent rescaling can appear at large endpoint times. The accumulated dissipative dose is
\begin{equation}
    \bar\gamma(T)T\sim \mu,
    \label{eq:constant_noise_dose_new}
\end{equation}
which is nearly independent of $T$. Once the system is beyond the Zeno-like transient and the dissipator acts only as a weak perturbation to the coherent dynamics, its leading effect is a nearly endpoint-independent damping of the relevant dynamical modes. For a macroscopic observable $O$, this motivates
\begin{equation}
    \langle O\rangle_{\rm noisy}(T)
    \simeq
    a_O(\mu)\langle O\rangle_{\rm ideal}(T) + c_{O}(\mu),
    \qquad
    T\gtrsim T_{\rm stat}.
    \label{eq:affine_rescaling_physical_new}
\end{equation}
The coefficients $a_O$ and $c_{O}$ depend on noise strength and circuit depth, but are approximately independent of the endpoint time in the stationary-rescaling window.

\begin{figure*}[ht]
\centering
\includegraphics[width=0.62\textwidth]{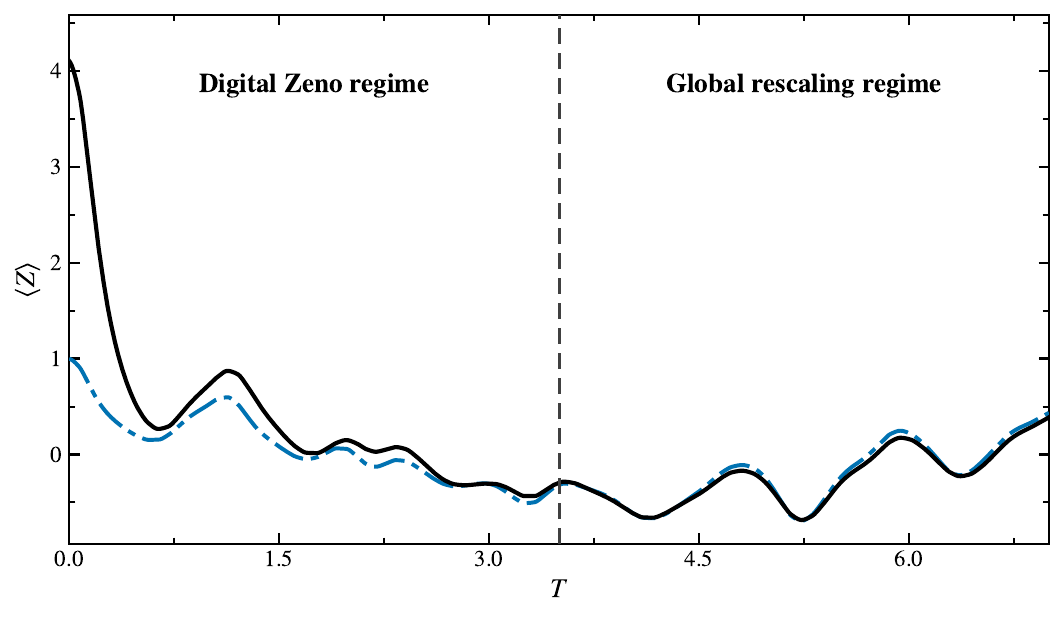}
\caption{Qualitative illustration of the two parts of a fixed-depth noisy time trace for a six-spin transverse-field Ising chain. The average magnetization is shown as a function of dimensionless time. The initial state is $|0\ldots0\rangle$, the chain is homogeneous with $J_i=J$ and $h_i=h$, and the plotted example uses $J=h$. The dashed line shows the ideal Trotterized evolution for $N=100$ layers. The solid line is the affinely reconstructed noisy signal at gate error probability $p=0.001$, not a physical magnetization curve. Its unphysical short-time values diagnose the failure of a stationary affine correction in the Zeno-like transient. The vertical dashed line marks the estimated crossover to the later observable-level stationary window.}
\label{fig:crossover_plot}
\end{figure*}

Figure~\ref{fig:crossover_plot} illustrates this operational separation for a small transverse-field Ising chain. The dashed line shows the ideal Trotterized evolution of the average spin magnetization $\langle Z \rangle =N_q^{-1}\sum_{j=1}^{N_q}\langle Z_j \rangle$. The solid line is not a directly measured physical magnetization; it is the noisy signal after a single affine reconstruction fitted in a later-time window. Its sharply unphysical short-time values are intentional in the presentation: they show that the stationary affine correction is being applied outside its domain of validity. The same fit becomes meaningful only after the crossover, where the channel-level stationary approximation has already formed for the observable considered.

The crossover trace in Fig.~\ref{fig:crossover_plot} was obtained from a density-matrix simulation of the first-order Trotterized transverse-field Ising model with periodic boundary conditions. We used the same ideal Trotter circuit, with $N=100$ layers, for both the noiseless and noisy data. Depolarizing noise was inserted after the elementary gates according to the model of errors we have introduced in Sec.~\ref{sec:protocol}. Shot noise was not included, so the figure isolates the deterministic bias produced by the circuit noise channel. The affine parameters were fitted in a later-time window and then extrapolated backward in time. The resulting unphysical short-time reconstruction is the intended diagnostic: it marks the regime in which a stationary affine projection is applied before the stationary-rescaling window has formed.

\section{Single-fault channels and Loschmidt echo}
\label{sec:single_fault}

We now derive the elementary building block of the global noise channel: the channel produced by one nontrivial fault pattern after averaging over the layer in which it occurs. The purpose of this section is specific. Rather than assuming a universal microscopic law for Loschmidt-echo decay, we construct a well-defined stationary single-fault channel and a physically meaningful time scale for its convergence.

\subsection{Fault layer representation}

A fault pattern within one layer is denoted by $\xi$. It is a particular combination of Pauli errors at the elementary noise locations of that layer. Its probability is denoted by $r_\xi$. For the depolarizing model of Eq.~\eqref{eq:one_qubit_depol},
\begin{equation}
    r_\xi=(1-p)^{n_{\rm loc}-|\xi|}\left(\frac{p}{3}\right)^{|\xi|},
    \label{eq:pattern_probability_new}
\end{equation}
where $|\xi|$ is the number of elementary Pauli faults in the pattern. The layer error probability is $q=\sum_{\xi\ne0}r_\xi$, and the conditional weight of a nontrivial pattern is
\begin{equation}
    w_\xi=\frac{r_\xi}{q}.
    \label{eq:w_xi_def}
\end{equation}

The faulty layer unitary is denoted by $U_{\Delta,\xi}$. For the Clifford plus Pauli-rotation Trotter layers considered here, every Pauli fault can be propagated to the beginning of the layer. This gives
\begin{equation}
    U_{\Delta,\xi}=U_\Delta V_{\Delta,\xi}P_\xi.
    \label{eq:fault_layer_factorization_new}
\end{equation}
Here $P_\xi$ is the propagated Pauli string and $V_{\Delta,\xi}$ is a coherent intra-layer correction caused by sign changes of small rotation generators during the propagation. Equivalently,
\begin{equation}
    P_\xi=U_{\Delta,\xi}\big|_{\Delta=0},
    \label{eq:P_xi_def_new}
\end{equation}
while
\begin{equation}
    V_{\Delta,\xi}=U_\Delta^\dagger U_{\Delta,\xi}P_\xi
    =\exp[-iv_\xi\Delta+O(\Delta^2)].
    \label{eq:V_xi_def_new}
\end{equation}
For local spin Hamiltonians and local propagated fault patterns, $v_\xi$ is local and has bounded norm. The leading stationary stochastic channel is obtained by setting $V_{\Delta,\xi}\to I$. This approximation and its limitations are discussed in Appendix~\ref{app:simplify_single_error}; the propagation rules and transpilation-dependent details are given in Appendix~\ref{app:transpilation_effects}.

The noisy channel of one Trotter layer is therefore
\begin{equation}
    \mathcal U_\Delta^{\mathcal N}
    =
    (1-q)\mathcal U_\Delta+
    \sum_{\xi\ne0} r_\xi\mathcal U_{\Delta,\xi},
    \label{eq:noisy_layer_sum}
\end{equation}
where $\mathcal U(\hat{\rho})=U\hat{\rho} U^\dagger$. Hereafter, the superscript $\mathcal{N}$ denotes the noisy implementation of an ideal unitary operation.

\subsection{Single-fault channels and memory time}

The full noisy circuit
\begin{equation}
    \mathcal Q^{\mathcal N}
    =
    \underbrace{\mathcal U_\Delta^{\mathcal N}\circ\cdots\circ\mathcal U_\Delta^{\mathcal N}}_{N},
    \label{eq:noisy_Q_channel}
\end{equation}
can be expanded in sectors with a fixed number of faulty layers by substituting \eqref{eq:noisy_layer_sum} directly into \eqref{eq:noisy_Q_channel}. We first isolate the one-fault sector. Suppose that a nontrivial fault pattern $\xi$ occurs in exactly one of the $N$ layers. After neglecting $V_{\Delta,\xi}$, summing over the possible noisy-layer positions, and passing to the continuous limit, we obtain the single-fault contribution $\mathcal{S}_{\xi}$ to the full quantum channel $\mathcal{Q}^{\mathcal{N}}$ of the Trotter circuit:
\begin{equation}
    \mathcal S_\xi(\hat{\rho})
    \simeq
    Nr_\xi(1-q)^{N-1}
    \mathcal U(T)\circ \mathcal M_\xi^{(T)}(\hat{\rho}),
    \label{eq:one_err_mean_pre_stationary}
\end{equation}
where $\mathcal U(T)(\hat{\rho})=e^{-iHT}\hat{\rho} e^{iHT}$ and
\begin{equation}
    \mathcal M_\xi^{(T)}(\hat{\rho})
    =
    \frac{1}{T}\int_0^T dt\,
    P_\xi(t)\hat{\rho} P_\xi(t).
    \label{eq:M_xi_t_new}
\end{equation}
Here
\begin{equation}
    P_\xi(t)=e^{iHt}P_\xi e^{-iHt}.
    \label{eq:P_xi_heisenberg_new}
\end{equation}
Equation~\eqref{eq:M_xi_t_new} is a random-unitary channel: it is an average of conjugations by the Hermitian unitary $P_\xi(t)$. The finite-$T$ expression naturally places the averaged fault before the final ideal evolution. We show that, for the systems and Trotter-layer transpilations considered here, the single-fault channels $\mathcal{M}_{\xi}^{(T)}$ have a stationary limit, $\mathcal{M}_{\xi}^{(T)} \rightarrow \mathcal{M}_{\xi}, T\rightarrow\infty$. Physically, convergence of the single-fault channels to their stationary limits corresponds to the loss of memory of the layer in which the error pattern $\xi$ occurred. The convergence rate can be related directly to the relaxation of the Loschmidt-echo amplitude. This relation can be understood qualitatively by rewriting Eq.~\eqref{eq:M_xi_t_new} after inserting the identity $P_{\xi}^2 = I$:
\begin{equation}
    \mathcal{M}_{\xi}^{(T)}(\hat{\rho}) = \frac{1}{T}P_{\xi}\int_{0}^{T}e^{i(H +\delta E)t}e^{-iHt}\hat{\rho}e^{iHt}e^{-i(H + \delta E)t}dtP_{\xi},
\end{equation}
where $\delta E = P_{\xi}HP_{\xi} - H$ is a local perturbation in our setting. We therefore identify the single-fault-pattern memory time $T^{*}_{\xi}$ with the crossover time of the running-averaged Loschmidt echo for the perturbation defined by $P_{
\xi}$. Statements about the convergence rate of $\mathcal{M}_{\xi}^{(T)}$ are proved in Appendix~\ref{app:channel_properties}. When a Fermi-golden-rule decay regime applies, $T_{\xi}^{*}$ can be estimated as the characteristic decay scale $\Gamma^{-1}$; for details of the possible Loschmidt-echo decay regimes, see \cite{Gorin_2006_LE}. We define $T^{*}$ as an effective maximum over the relevant values of $T_{\xi}^{*}$. We do not provide a universal macroscopic formula for $T^{*}_{\xi}$; its value depends on the Hamiltonian, the perturbation, symmetries, and finite-size recurrences. For the transverse-field Ising model, useful analytical and numerical intuition is available from studies of Loschmidt echoes and dynamical phase transitions \cite{Polkovnkiov_dynamical_phase_transitions, Silva_ising_echo}.

We emphasize that the convergence toward the
stationary limit $\mathcal{M}_{\xi}^{(T)} \rightarrow \mathcal{M}_{\xi}$ takes place only in terms of the average over most of the possible bulk fault positions, with edge effects controlled by the same Loschmidt echo relaxation, as we show in Appendix~\ref{app:channel_properties}. In this stationary limit, for sufficiently large $T$, the channel commutes with the ideal dynamics and may therefore be moved to the end of the circuit:

\begin{equation}
    \mathcal M_\xi\circ\mathcal U(T)
    =
    \mathcal U(T)\circ\mathcal M_\xi.
    \label{eq:time_translation_invariance_main}
\end{equation}
This is the key property that permits the separation of ideal dynamics and stationary noise. The basis for this time invariance with a more detailed discussion of the finite-time channel commutator norm is also given in Appendix~\ref{app:channel_properties}. Accordingly, in the stationary regime,
\begin{equation}
    \mathcal S_\xi(\hat{\rho})
    \simeq
    Nr_\xi(1-q)^{N-1}
    \mathcal M_\xi\circ\mathcal U(T)(\hat{\rho}).
    \label{eq:one_err_mean}
\end{equation}

The memory-time interpretation is crucial for the fixed-depth protocol. A circuit with many faults forms a stationary global noise channel only when the endpoint time is large enough for typical fault insertions to self-average. This is the origin of the sufficient conservative time scale $T_{\rm stat}\sim NqT^\ast$ derived next.

\section{Binomial global noise channel and Poisson limit}
\label{sec:poisson}

We now extend the single-fault result to a finite-depth noisy circuit. The number $M$ of faulty layers is binomially distributed with layer-fault probability $q$. It is therefore useful to keep the binomial structure as the primary result and to regard the Poisson exponential as its dilute-layer limit.

For a fixed number $M$ of faulty layers, we sum over all ordered layer positions and over all nontrivial fault patterns. If the endpoint time is large compared with the memory time required by each insertion,
\begin{equation}
    T\gtrsim M T^\ast,
    \label{eq:t_larger_M_Tstar_new}
\end{equation}
then each insertion can be replaced by its stationary averaged channel $\mathcal{M}_{\xi}$. In this derivation, the temporal ordering of the error insertions is essential, because the propagated error operators $P_{\xi_{m}}(t_{m})$ corresponding to different error positions and patterns do not generally commute. Equation~\eqref{eq:t_larger_M_Tstar_new} therefore yields a conservative lower bound on the simulation time required for the stationary approximation to apply. If, however, the dominant error patterns generate propagated operators that approximately commute over a substantial portion of the evolution, the ordering constraint can be weakened, and the factor $M$ in the corresponding scaling may be reduced or eliminated. In this case, the error contributions can be effectively averaged over the full simulated time interval rather than over $M$ separately ordered intervals. For local Hamiltonians and local error insertions, the error introduced by such reorderings can be estimated using Lieb--Robinson bounds, which control commutators of spatially separated Heisenberg-evolved operators~\cite{LiebRobinson1972,NachtergaeleSims2006,KlieschGogolinEisert2014}. A complete removal of the factor $M$, however, requires the accumulated commutator contribution over all relevant pairs of error insertions to remain negligible. In this work, we consider the worst case. The fixed-$M$ contribution then factorizes as
\begin{equation}
    \mathcal S_M^{\rm bin}(T)
    \simeq
    \binom{N}{M}q^M(1-q)^{N-M}
    \overline{\mathcal M}^{\,M}\circ\mathcal U(T),
    \label{eq:SM_main_new}
\end{equation}
where
\begin{equation}
    \overline{\mathcal M}=\sum_{\xi\ne0}w_\xi\mathcal M_\xi
    \label{eq:Mbar_new}
\end{equation}
is the stationary single-fault channel averaged over nontrivial patterns inside one faulty layer. A derivation from ordered sums is given in Appendix~\ref{app:error_summ}.

Summing Eq.~\eqref{eq:SM_main_new} over $M$ gives the stationary global noise channel
\begin{equation}
    \mathcal Q^{\mathcal N}(T)
    \simeq
    \mathcal E_{N,q}^{\rm bin}\circ\mathcal U(T),
    \qquad
    \mathcal E_{N,q}^{\rm bin}
    =
    \left[(1-q)\mathcal I+q\overline{\mathcal M}\right]^N.
    \label{eq:final_channel}
\end{equation}
This is the central channel-level result. The endpoint-time dependence is carried by the ideal evolution $\mathcal U(T)$, while the leading noise channel depends on the fixed circuit depth and the microscopic error model through $N$, $q$, and the stationary single-fault average $\overline{\mathcal M}$.

The Poissonian expression is recovered from Eq.~\eqref{eq:final_channel} in the dilute-layer limit
\begin{equation}
    q\to0,
    \qquad
    Nq=\mu \quad \text{fixed},
    \label{eq:poisson_limit_conditions}
\end{equation}
namely
\begin{equation}
    \mathcal E_{N,q}^{\rm bin}
    \longrightarrow
    \mathcal E_\mu^{\rm Pois}
    =
    \exp\left[\mu(\overline{\mathcal M}-\mathcal I)\right].
    \label{eq:poisson_channel}
\end{equation}
For finite $q$, the binomial channel should be used. On an eigenmode of $\overline{\mathcal M}$ with eigenvalue $\eta$, the binomial damping factor is $[1-q(1-\eta)]^N$, whereas the Poisson approximation gives $\exp[-\mu(1-\eta)]$. The two agree to leading order when the corrections of order $\mu q(1-\eta)^2$ are negligible.

The conservative condition for the formation of the stationary channel is obtained by applying Eq.~\eqref{eq:t_larger_M_Tstar_new} to the values of $M$ that carry most of the binomial weight. A high-probability estimate gives
\begin{equation}
    M_\delta
    =
    Nq+O\!\left(\sqrt{Nq(1-q)\ln(1/\delta)}\right) \sim Nq\left(1 + O\left(\sqrt{\frac{n_{O}}{n_{\rm loc}}}\right)\right),
    \label{eq:M_delta_main}
\end{equation}
so that
\begin{equation}
    T_{\rm stat}(\delta)
    \sim
    M_\delta T^\ast,
    \qquad
    T_{\rm stat}\sim NqT^\ast=\mu T^\ast
    \label{eq:Tstat_main_new}
\end{equation}
at leading order. Here $n_{O}$ is an observable-dependent damping scale related to its Pauli weight, while $\delta$ specifies the required accuracy of the total channel and may be tied to the depolarizing damping factor $\sim (1 - p)^{n_{O}N}$. Below this scale, fault insertions are not guaranteed to self-average, and the noise channel can remain strongly endpoint dependent. This is the channel-level expression of the short-time Zeno-like distortion.

Appendix~\ref{app:poisson_lim} summarizes the finite-depth corrections, the Chernoff estimate used in Eq.~\eqref{eq:M_delta_main}, and the additional dilute-layer requirements under which the ordered-sum continuum approximation and the Poisson limit are controlled.

\section{Spectral concentration and observable-level error mitigation}
\label{sec:rescaling}

The factorized channel in Eq.~\eqref{eq:final_channel} is useful because it converts a microscopic time-dependent noise process into a stationary correction acting after the ideal evolution. This is the main practical advantage of the fixed-depth protocol. Beyond the crossover, if the $\mathcal{E}^{\rm bin}_{N,q}$ channel is approximately depolarizing, the leading noise contribution need not be learned independently at every endpoint time; it can be calibrated once and applied to an extended part of the dynamical curve.

In the Heisenberg picture, for an initial density operator $\hat{\rho}_{0}$:
\begin{equation}
    \langle O\rangle_{\rm noisy}(T)
    =
    \operatorname{Tr}\left[
    \hat{\rho}_0\,\mathcal U^\dagger(T)
    \left(\mathcal E_{N,q}^{\rm bin}(O)\right)
    \right],
    \label{eq:observable_heisenberg_new}
\end{equation}
where the Poissonian expression $\mathcal E_\mu^{\rm Pois}$ may be substituted for $\mathcal E_{N,q}^{\rm bin}$ only in the dilute-layer limit of Eq.~\eqref{eq:poisson_limit_conditions}.

Equation~\eqref{eq:observable_heisenberg_new} is where the distinction between channel-level stationarity and observable-level depolarization becomes important. If $\mathcal E_{N,q}^{\rm bin}$ were completely depolarizing on the full operator algebra, all traceless observables would be suppressed in the same way. The channel derived above is more structured: at moderate or higher noise levels $\mu$, different operator modes may have different damping factors and may mix under $\mathcal E_{N,q}^{\rm bin}$. We now describe how the total channel in Eq.~\eqref{eq:final_channel} changes with the noise level $\mu$ and identify when stationary rescaling is justified. For low-to-moderate noise in deep circuits, the Poissonian limit \eqref{eq:poisson_channel} may be used when the dilute-layer conditions hold. If $\mu$ is sufficiently small, the decay rates of typical operator modes are close to one another. At the observable level, this is equivalent to approximating Eq.~\eqref{eq:poisson_channel} by a global depolarizing channel with fidelity $F = (1 - q)^{N} \approx e^{-\mu}$ on the nonzero-Bohr-frequency, time-dependent sector of operator space. The spectral estimates derived in Appendix~\ref{app:M_final_prop} give the following condition for this simplification:

\begin{equation}
    \mu^{2}\left(\sum_{P}\widetilde{w}^{2}_{P}\right) \approx \mu^{2}\left(\sum_{\xi, \zeta}w_{\xi}w_{\zeta}|\mathrm{Tr}\left(P_{\xi}P_{\zeta}\right)|^{2}/d^{2}\right) \lesssim 1,
    \label{eq:global_depol_req}
\end{equation}
For the spectrum of $\overline{\mathcal{M}}$, $P$ labels a distinct propagated Pauli error and $\widetilde{w}_{P}$ is its conditional weight, defined as $\widetilde{w}_{P} = \sum_{P_{\xi} = \pm P} w_{\xi}$. In the one-error-per-layer limit, $\sum_{P}\widetilde{w}^{2}_{P} \sim 1/n_{\rm loc}$. At the same level of approximation, the normalized Frobenius distance between the global depolarizing channel and $\mathcal{E}^{\rm bin}_{N, q}$ obeys the corresponding estimate

\begin{equation}
    ||\mathcal{E}^{\rm bin}_{N, q} - \mathcal{D}_{(1-q)^{N}}||_{2}^{2} \leq \mu^{2}\left(\sum_{P}\widetilde{w}^{2}_{P}\right),
    \label{eq:distance_depol}
\end{equation}

with the same estimate holding for the Poisson limit. Thus, when a layer contains many distinct fault locations, low-to-moderate noise can produce an effectively depolarizing action on the observables. This regime provides the theoretical basis for the stationary observable rescaling seen in the simulations. Under these conditions, the action of the noise channel $\mathcal{E}^{\rm bin}_{N,q}$ on the observables of interest can be approximated by a simple affine transformation:
\begin{equation}
    \mathcal E_{N,q}^{\rm bin}(O)
    \simeq
    a_O(N,q)(O - O_{\rm stat}) + b_{O}(N,q)O_{\rm stat},
    \label{eq:affine_operator_new}
\end{equation}
where $O_{\rm stat}$ denotes the zero-Bohr-frequency (energy-block-diagonal) projection of the observable. Thus, $O_{\rm stat}$ does not produce any time dependence. Consequently,
\begin{equation}
    \langle O\rangle_{\rm noisy}(T)
    \simeq
    a_O(N,q)\langle O\rangle_{\rm ideal}(T) + c_{O}(N, q),
    \qquad T\gtrsim T_{\rm stat}.
    \label{eq:observable_rescaling_new}
\end{equation}
The factor $a_O$ should be interpreted as an observable contrast factor rather than as a channel fidelity, while $c_{O}$ is a constant offset that can generally be nonzero. Operationally, however, $a_{O}$ plays a similar role in estimating a single observable: if the ideal value is reconstructed by dividing by $a_O$, the sampling cost of the nontrivial dynamical part increases approximately as $|a_O|^{-2}$. Equation~\eqref{eq:global_depol_req} also predicts the breakdown of the stationary-rescaling picture when error rates are strongly imbalanced across different operations on the quantum device. In the worst-case scenario, where only a single operation is severely noisy, one has $\sum_{P}\widetilde{w}_{P}^{2}\sim O(1)$. At noise levels $\mu$ beyond the regime specified by Eq.~\eqref{eq:global_depol_req}, the noise dependence of the observable typically becomes nontrivial, and a simple rescaling may no longer recover the qualitative dynamics. In this situation, one may instead use multiexponential ZNE to reconstruct the observable. For global-folding, the circuit is replaced by
\begin{equation}
    \mathcal Q^{\rm \mathcal{N}, folded}_n
    =
    \mathcal Q^{\rm \mathcal{N}}
    \circ
    \left[(\mathcal Q^{\dagger})^{\rm \mathcal{N}}\circ\mathcal Q^{\rm \mathcal{N}}\right]^n,
    \label{eq:global_folding_new}
\end{equation}
where $(\mathcal{Q}^{\dagger})^{\mathcal{N}}$ denotes the same Trotter circuit implementing inverse unitary evolution, obtained by reversing the signs of the unitary-gate rotation angles within each layer. The noise channel of $(\mathcal{Q}^{\dagger})^{\mathcal{N}}$ is the same $\mathcal{E}_{N, q}^{\rm bin}$: the distribution of error patterns $P_{\xi}$ does not change because it is fixed by the transpilation, and the stationary limit in Eq.~\eqref{eq:M_xi_t_new} is invariant under time reversal. Since the stationary channel commutes with the ideal evolution, we obtain
\begin{equation}
    \mathcal Q^{\rm \mathcal{N}, folded}_n
    \simeq
    \left(\mathcal E_{N,q}^{\rm bin}\right)^{2n+1}\circ\mathcal U.
    \label{eq:global_folding_factor_new}
\end{equation}
In the dilute-layer limit, this becomes $(\mathcal E_\mu^{\rm Pois})^{2n+1}\circ\mathcal U$. Thus, global-folding amplifies the same stationary global noise channel by the expected factor $2n+1$, providing a microscopic justification for multiexponential global-folding zero-noise extrapolation \cite{Cai_2021_ZNE, ZNE_intro}. Similar reasoning applies to local folding when the folding operation changes the effective layer error probability without changing the relevant set of propagated error insertions.

The stationary exponential channel and the resulting multiexponential ZNE construction are justified only after the time $T_{\rm stat}$. This raises an important question: how does the mitigation procedure change across the stationary-channel crossover? Cai's general argument for multiexponential ZNE \cite{Cai_2021_ZNE} still applies in the Zeno-like regime before $T_{\rm stat}$, but the fit coefficients can depend strongly on the dynamics. This dependence effectively changes the number of modes, and hence the fit complexity, required for an accurate reconstruction of the observable. One should therefore expect a qualitative change in the ZNE procedure across the crossover. We illustrate this change by simulating multiexponential ZNE in a standard global-folding paradigm for the mean magnetization at a fixed noise
level and for different numbers of qubits in the transverse-field Ising Trotter circuits discussed above. The results show that $T_{\rm stat}$ separates the time axis $T$ into two error mitigation complexity regimes; see Fig.~\ref{fig:zne_six_panel_summary}. Before the $T_{\rm stat}$ crossover, a single-exponential approximation is insufficient and may even introduce an apparent coherent-like systematic error. Therefore, at least two modes are required for effective noise mitigation in this regime. Beyond the crossover, however, the simplest single-mode approximation becomes accurate and does not produce any discernible systematic error.

\newcommand{\panelabel}[1]{%
    \raisebox{1.0em}[0pt][0pt]{\textbf{(#1)}}\\[-0.3em]
}

\begin{figure*}[t]
    \centering

    \begin{minipage}[t]{0.485\textwidth}
        \centering

        \panelabel{a}
        \includegraphics[width=\linewidth]{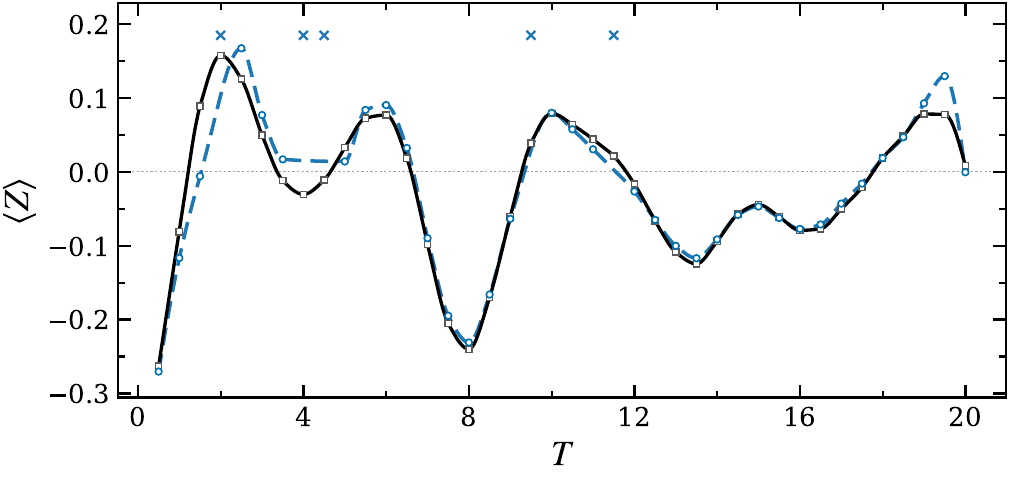}
        \vspace{0.5em}

        \panelabel{b}
        \includegraphics[width=\linewidth]{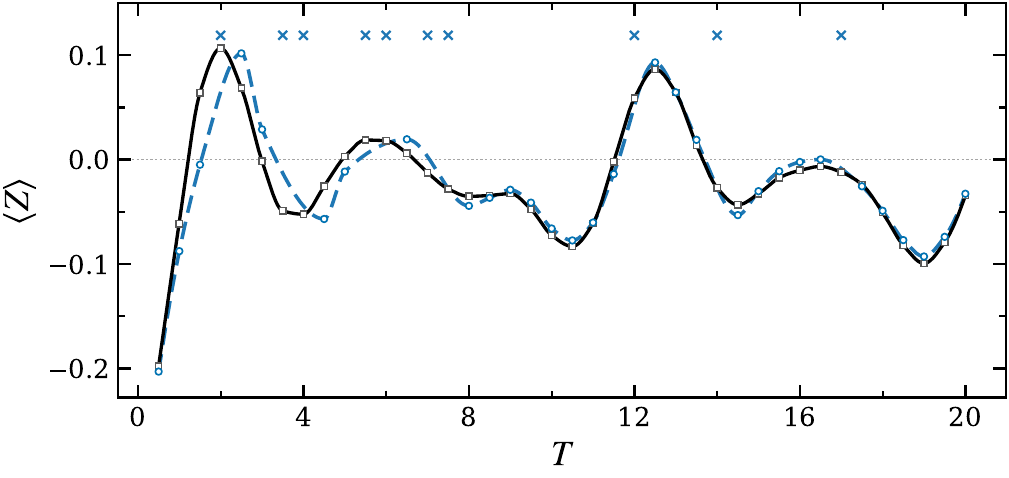}
        \vspace{0.5em}

        \panelabel{c}
        \includegraphics[width=\linewidth]{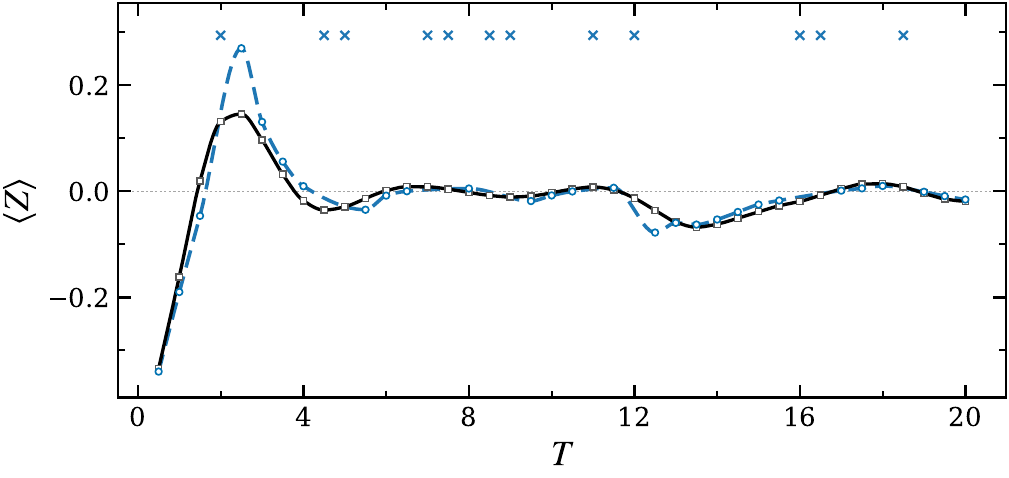}
    \end{minipage}
    \hfill
    \begin{minipage}[t]{0.485\textwidth}
        \centering

        \panelabel{d}
        \includegraphics[width=\linewidth]{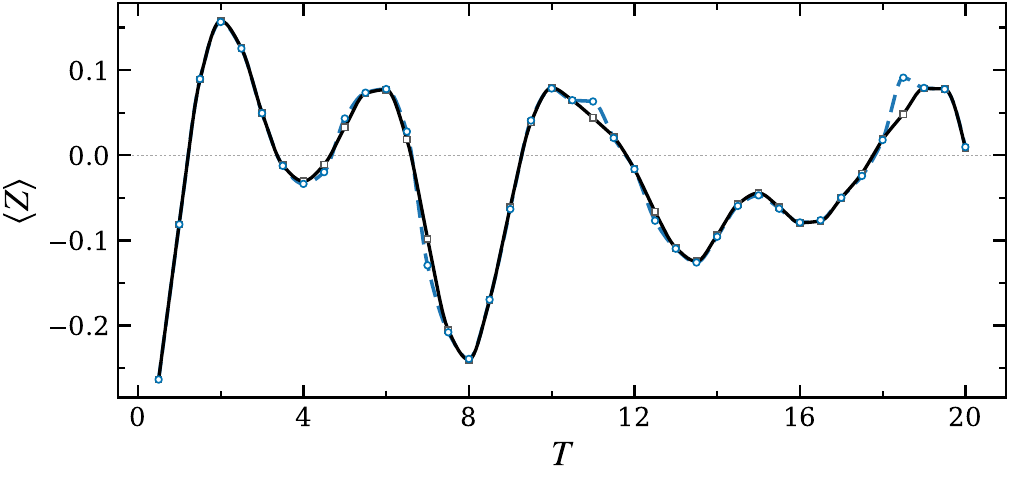}
        \vspace{0.5em}

        \panelabel{e}
        \includegraphics[width=\linewidth]{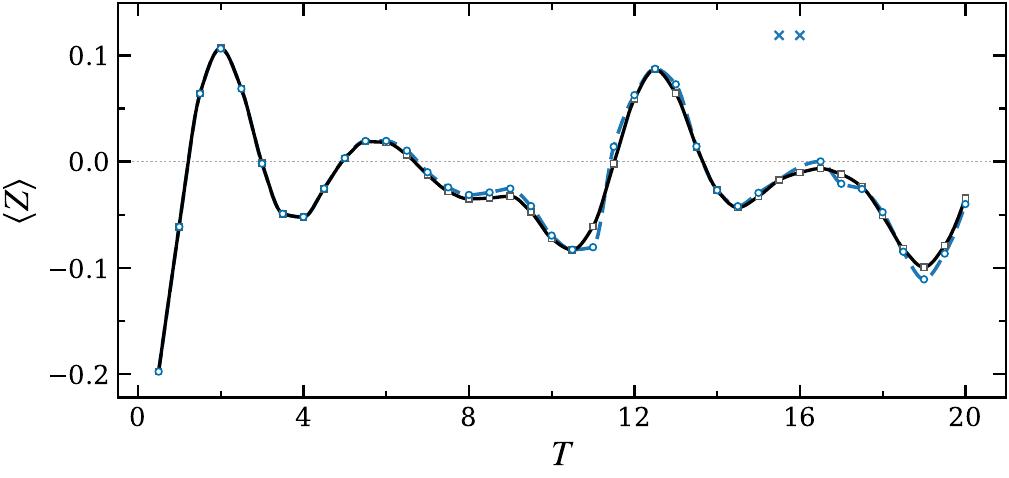}
        \vspace{0.5em}

        \panelabel{f}
        \includegraphics[width=\linewidth]{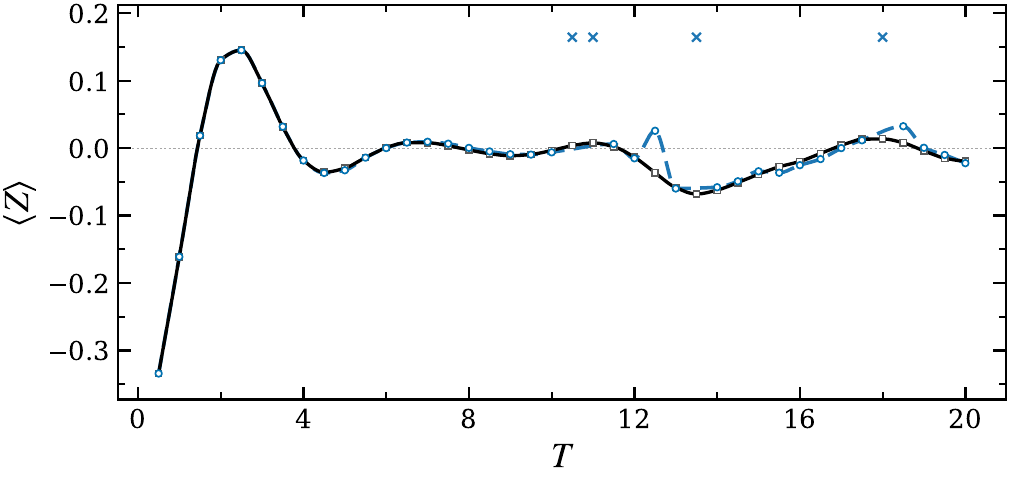}
    \end{minipage}

    \caption{Results of a multiexponential ZNE procedure for the numerically simulated mean magnetization of 6-, 8-, and 10-qubit transverse-field Ising chains with $J = 4h$ and corresponding initial states $|101101\rangle, |10110110\rangle, |1011011011\rangle$. The dynamics is simulated using a Trotter circuit of depth $N = 70$ with an error probability $p = 0.001$ per noise location. The first column, (a)--(c), corresponds to a single-exponential approximation, whereas the second column, (d)--(f), uses an optimally chosen number of exponents for the same simulation task. At each time point, we increase the number of modes until obtaining the first nondivergent extrapolation result consistent with the range of the true observable, using up to 4 modes in the provided example. We then vary the maximum ZNE global-folding factor $M_{\rm fold}$ and select the result with the fewest divergent points. Panels (a,d), (b,e), and (c,f) correspond to 6, 8, and 10 qubits, respectively. The solid line depicts the ideal, noiseless simulation result, whereas the dashed data shows the ZNE-reconstructed data obtained from the same Trotter circuit, folded up to $M_{\rm fold}$ times. Crosses denote data points where ZNE fails, producing an unphysical value of the considered observable. A complexity crossover in time, associated with $T_{\rm stat}$, marks the time range beyond which a single-exponential fit becomes sufficient for accurate reconstruction. Before this crossover, higher-order approximations are essential for an accurate fit. The figure therefore illustrates this transition between ZNE complexity regimes controlled by the internal simulation time $T$.
    }
    \label{fig:zne_six_panel_summary}
\end{figure*}

We finally comment on the strong-noise regime. Here the binomial channel representation is the appropriate starting point. For fixed $q$ and large depth, the binomial weight in Eq.~\eqref{eq:final_channel} is concentrated near its mean number of faulty layers, so the dominant terms involve powers close to $\overline{\mathcal{M}}^{\mu}$. Employing the spectral representation of the action of $\overline{\mathcal{M}}^{\mu}$ on the observable $O$, one expects sufficiently strong noise to select the slowest nontrivial mode:

\begin{equation}
    \mathcal E_{N,q}^{\rm bin}\left(O\right) \sim (1 - q + q\lambda_{2})^{N}O_{\lambda_{2}},
\end{equation}
where $O_{\lambda_{2}}$ denotes an eigenoperator, or the projection onto the corresponding degenerate eigenspace, associated with the largest nontrivial eigenvalue of the $\overline{\mathcal{M}}$ channel; the identity mode is excluded. Because $\overline{\mathcal{M}}$ commutes with $\mathcal{U}(T)$, and by the usual open-system intuition that coherences decay faster than populations relax toward a thermal Gibbs state, we expect $O_{\lambda_{2}}$ to be diagonal in the $H$ basis in a typical situation. In this case, $O_{\lambda_{2}}$ is the slowest-decaying part of the observable. Under this assumption, at strong noise and beyond the $T_{\rm stat}$ scale, the time-dependent part of the observable is damped away, leaving a constant offset associated with $O_{\lambda_{2}}$. If the Markovian dynamics has a long-lived coherence sector, however, $O_{\lambda_{2}}$ may be nondiagonal, and a nonvanishing oscillatory mode can survive even under strong noise. Appendix~\ref{app:M_final_prop} gives the theoretical basis for these statements about the different regimes of the $\mathcal{E}_{N, q}^{\rm bin}, \mathcal{E}_{\mu}^{\rm Pois}$ channels.

\begin{figure*}[!h]
    \centering
    \begin{minipage}{0.48\textwidth}
        \centering
        \begin{overpic}[width=\linewidth]{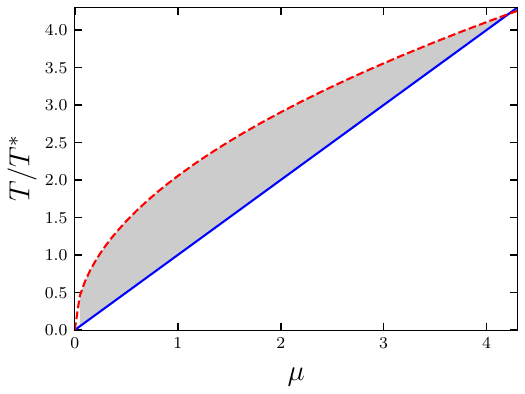}
            \put(20,65){\large \text{(a)}}
        \end{overpic}
    \end{minipage}
    \hfill
    \begin{minipage}{0.49\textwidth}
        \centering
        \begin{overpic}[width=\linewidth]{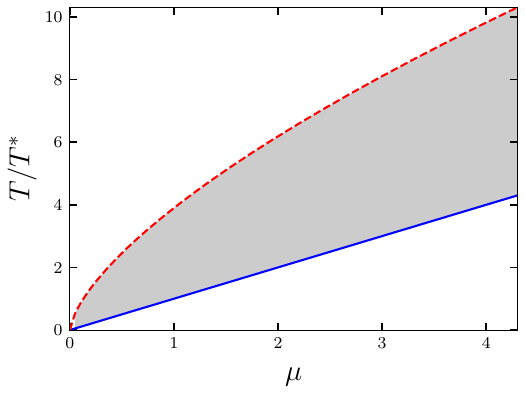}
            \put(20,65){\large \text{(b)}}
        \end{overpic}
    \end{minipage}
    \caption{Illustrative conservative windows for the stationary-channel description in first- and second-order Trotter circuits, (a) and (b), respectively. The axes show the mean number of faulty layers $\mu$ and normalized endpoint time $T/T^\ast$. The solid lines represent the conservative self-averaging condition $T>\mu T^\ast$, while the dashed lines show the upper bound imposed by the Trotterization-error tolerance. The highlighted regions between these lines are therefore consistency regions for the channel-level approximation; the affine projection for a given observable is an additional observable-level condition. The plotted examples correspond to the transverse-field Ising model with parameters $N_{q}=6$, $J=h$, $\alpha=0.1$, and gate error level $p=0.001$.}
    \label{fig:rescaling_region}
\end{figure*}

The conservative window in which the stationary channel and its consequences apply is constrained by several requirements. First, the endpoint time must exceed the worst-case self-averaging scale. Second, the coherent product-formula error must remain below the desired tolerance. A compact leading-order set of conditions is
\begin{equation}
    \begin{gathered}
        T > \mu T^{*},\\
        \frac{T^{g+1}}{N^{g}} < \alpha,
    \end{gathered}
    \label{eq:full_region_system_basic}
\end{equation}
where $g$ is the product-formula order and $\alpha$ is the chosen Trotter-error tolerance. Using $N=\mu/q$, the Trotter bound may also be written as
\begin{equation}
    T < \alpha^{\frac{1}{g+1}}
        \left(\frac{\mu}{q}\right)^{\frac{g}{g+1}}.
    \label{eq:full_region_system}
\end{equation}
A more careful high-probability version replaces the first inequality by Eq.~\eqref{eq:Tstat_main_new}. Representative examples for the first- and second-order Trotterization schemes are shown in Fig.~\ref{fig:rescaling_region}; the plotted regions should be interpreted as necessary stationary-channel windows rather than as a proof of the affine projection for every observable.

The window plot in Fig.~\ref{fig:rescaling_region} combines the leading self-averaging requirement $T\gtrsim \mu T^{\ast}$ with a product-formula accuracy bound. It should be read as a conservative consistency window for applying the stationary-channel approximation. Whether a particular observable is then accurately described by an affine contrast correction is checked at the observable level, for example by residuals or independent calibration data. This figure shows that Trotter decompositions of higher order $g$ are advantageous because they broaden the consistency window. In general, however, one should account for a possible increase in the noise level $q$, which depends on the available transpilation \cite{Areg}.

Equation~\eqref{eq:observable_rescaling_new} is the error mitigation content of the stationary-channel result. The coefficients $a_O$ and $c_{O}$ may be obtained from calibration circuits, from a classically tractable part of the evolution, or from a zero-noise-extrapolation fit. Once calibrated, the same coefficients can be used over the endpoint-time window in which the stationary-rescaling approximation holds. In this sense, the fixed-depth protocol converts a complicated hardware-noise problem into a low-dimensional calibration problem. Further numerical demonstrations for the MFIM and XX
models, covering integrable and nonintegrable regimes and several observables, are presented in Appendix ~\ref{app:numerical_sim}.

The rescaling procedure is also naturally compatible with other Trotter protocols. Fixed-step or fixed-error simulations may provide accurate short-time data, Trotter-convergence checks, or anchor points for the calibration. Fixed-depth simulations, on the other hand, are designed to produce a time trace with a stable leading noise channel. The protocols therefore probe the same target dynamics from different directions. Their agreement after rescaling gives a stringent consistency check, while their differences can reveal the boundary of the stationary-rescaling window.

Since we consider the fixed-depth protocol as a valuable alternative to fixed-error and fixed-step approaches, providing a significant improvement to the exponential overhead bottleneck of standard error mitigation techniques, it is important to evaluate the overall gate complexity $G$ and compare it with conventional approaches. We define $G$ as the total number of gates required to obtain $K$ equally spaced time points at which the observable is measured up to $T_{\rm max}$ while maintaining the same maximum tolerated Trotter
error determined by the maximum circuit depth $N_{\rm max}$. We estimate
the complexity of the fixed-depth protocol for the optimistic case in which one does not need to implement complex error mitigation schemes such as $\rm ZNE$ and $\rm PEC$ to mitigate noise. It will be assumed that the relation \eqref{eq:observable_rescaling_new} is used, with the same parameters for most of the $K$ data points. Here, the cost of estimating $a_{O}$ and $c_{O}$ is neglected, which is justified for sufficiently long simulations.

Thus, for the fixed-depth case $(FD)$:

\begin{equation}
    G_{FD} \simeq n_{\rm loc}KN_{\rm max}\exp(2qN_{\rm max})\epsilon^{-2},
    \label{eq: fixed_depth_complexity}
\end{equation}
where $n_{\rm loc}N_{\rm max}$ is the total number of gates needed for one experiment and $\exp(2qN_{\rm max})\epsilon^{-2}$ represents the $F^{-2}\epsilon^{-2}$ fidelity overhead needed to estimate the
observable with relative accuracy $\epsilon$. Hereafter, we omit the common relative-precision factor $\epsilon^{-2}$, since it can be absorbed into
an overall prefactor $n_{\rm loc}K$ and will not affect our qualitative comparison of different complexities.

To estimate the complexity of error mitigation used in conventional protocols, we will consider a simplified model of the $\rm ZNE$ procedure, defined via global-folding circuit structure \eqref{eq:global_folding_new} with a fixed maximal $M_{\rm fold}$ folding factor for each of $K$ data points. For fixed-step and fixed-error protocols $(P = FS, FE)$, we can write a single expression that accounts for the $M_{\rm fold}-\rm ZNE$ error mitigation procedure analogous to \eqref{eq: fixed_depth_complexity} applied to each circuit:

\begin{equation}
    \begin{gathered}
    G_{P} \simeq n_{\rm loc}\sum_{n = 0}^{K}\sum_{m = 0}^{M_{\rm fold}}(2m + 1)N_{P}(n)\exp\left(2q\left(2m + 1\right)N_{P}(n)\right) = \\
    \frac{n_{\rm loc}}{2}\frac{\partial}{\partial q}\left\{\sum_{n = 0}^{K}\exp\left(2qN_{P}(n)\right)\frac{1 - \exp(4q(M_{\rm fold} + 1)N_{P}(n))}{1 - \exp(4qN_{P}(n))}\right\},\\
    \end{gathered}
    \label{eq: FD and FS complexity}
\end{equation}
where $m$ stands for the folding step of the $\rm ZNE$ and  $N_{P}(n)$ denotes the number of Trotter layers used in the corresponding protocol at the $n$-th time point. After simplifying the expression and approximating the sum by an integral for $K \gg 1$, \eqref{eq: FD and FS complexity} turns into:
\begin{equation}
    \begin{gathered}
        G_{P} \simeq\frac{n_{\rm loc}}{2}\frac{\partial}{\partial q}\left\{\int_{0}^{K}\exp\left(2q\left(M_{\rm fold} + 1\right)N_{P}(x)\right) \frac{\sinh\left(2q\left(M_{\rm fold} + 1\right)N_{P}(x)\right)}{\sinh\left(2qN_{P}(x)\right)}dx\right\}, \\
    N_{FS}(x) = N_{\rm max}\frac{x}{K},\quad N_{FE}(x) = N_{\rm max}\left(\frac{x}{K}\right)^{\frac{g + 1}{g}},
    \end{gathered}
    \label{eq: FD and FS simplified complexity}
\end{equation}
One can note that expression \eqref{eq: FD and FS simplified complexity} can be written purely in terms of the overall noise parameter $\mu' = 2qN_{\rm max}$. For the remainder of the complexity analysis,
we use the parameter $\mu'$ instead of $q$.



Now, we are interested in examining two key limits: the low-noise regime, characterized by $\mu' \ll 1$, and the NISQ-oriented regime of substantial noise, characterized by $\mu' \gtrsim 1$, in order to understand the dependence of the relative complexity on the available device noise parameters. For the small-noise regime, expanding \eqref{eq: FD and FS simplified complexity}, we obtain:

\begin{equation}
    \begin{gathered}
    G_{FS}^{(\mu' \rightarrow0)} \approx
    n_{\mathrm{loc}}K N_{\rm \max}
    \left[
    \frac{(M_{\rm fold}+1)^2}{2}
    +
    \frac{\mu' (M_{\rm fold}+1)(2M_{\rm fold}+1)(2M_{\rm fold}+3)}{9}
    \right], \\
    G^{(\mu' \rightarrow 0)}_{FE}
    \approx
    n_{\rm loc}K N_{\max}
    \left[
    \frac{g(M_{\rm fold}+1)^2}{2g+1}
    +
    \frac{g\,\mu' (M_{\rm fold}+1)(2M_{\rm fold}+1)(2M_{\rm fold}+3)}
    {3(3g+2)}
    \right].
    \end{gathered}
    \label{eq: small_noise_complexity}
\end{equation}

For the substantial noise case, we approximately obtain:

\begin{equation}
\begin{gathered}
G_{FS}
\simeq
n_{\rm loc} K N_{\rm max}
\frac{
\exp\left((2M_{\rm fold}+1)\mu'\right)
}{\mu'}
\left(
1-\frac{1}{(2M_{\rm fold}+1)\mu'}
\right), \\
G_{FE}
\simeq
\frac{g}{g + 1}G_{FS}.
\end{gathered}
\label{eq: big_noise_complexity}
\end{equation}

Now, we can examine both limits with respect to $G_{FD}$. In the
regime close to fault tolerance we obtain:

\begin{equation}
    \frac{G^{(\mu' \rightarrow 0)}_{FD}}{G^{(\mu' \rightarrow 0)}_{FS}} \sim 2, \quad \frac{G^{(\mu' \rightarrow 0)}_{FD}}{G^{(\mu' \rightarrow 0)}_{FE}} \sim 2 +\frac{1}{g}.
    \label{eq: small_noise_compare}
\end{equation}
Since in this regime the error mitigation procedure can be omitted for the $FS$ and $FE$ protocols, the fixed-depth protocol introduces a gate overhead of up to a factor of $3$. However, in the NISQ regime set by $\mu' \gg 1$:

\begin{equation}
    \frac{G_{FD}}{G_{FS}} \sim \mu'\exp(-2M_{\rm fold}\mu'), \quad \frac{G_{FD}}{G_{FE}} \sim \left(1 + \frac{1}{g}\right)\mu'\exp\left(-2M_{\rm fold}\mu'\right).
    \label{eq: big_noise_compare}
\end{equation}

We see that when the noise effect is small, the fixed-depth protocol naturally incurs a substantial gate overhead. But as the noise level increases, the fixed-depth protocol can provide, in the optimistic case where the affine transformation coefficients are estimated either classically or via a short-time experiment, an exponential gate advantage compared with conventional approaches. This should be viewed as a potential advantage whose applicability must be verified for each specific regime, since at sufficiently high noise levels the condition \eqref{eq:global_depol_req} prevents the use of simple observable rescaling, and the exponential advantage regime becomes inaccessible.

\section{Conclusion}
\label{sec:conclusion}

We have analyzed noisy fixed-depth Trotter simulations of many-body dynamics. The central point is that the fixed-depth protocol allows one to structure noise. The number of layers is chosen from the Trotter-accuracy requirement at the largest endpoint time and is then kept fixed over the interval of interest. This choice is not noise-optimal for every short-time data point, but it keeps the total hardware-noise dose approximately independent of endpoint time. The benefit is a much simpler leading noise structure.

This channel structure also explains the operational Zeno-like crossover. At short endpoint times the coherent rotation angle per layer is small compared with the layer error dose, and stochastic faults repeatedly randomize the coherences needed for the target unitary motion. At larger endpoint times, after local faults have self-averaged under the target dynamics, the coherent evolution determines the shape of the signal while the accumulated noise dose remains nearly fixed. The resulting channel need not be completely depolarizing. The useful effect is more selective: for chosen observables and state families, the channel can act as an almost time-independent contrast loss, i.e., as observable-level depolarization.

Starting from a microscopic depolarizing error model, we represented local Pauli faults as propagated Pauli insertions whose positions are averaged over the ideal dynamics. The convergence of the corresponding single-fault channel defines a memory time $T^\ast$ and is naturally diagnosed by a Loschmidt echo of the local perturbation $\delta E=P_\xi H P_\xi-H$. For independent faulty layers, the finite-depth stationary global channel is binomial,
\begin{equation}
    \mathcal Q^{\mathcal{N}}(T)
    \simeq
    \left[(1-q)\mathcal I+q\overline{\mathcal M}\right]^N
    \circ\mathcal U(T).
\end{equation}
The Poissonian exponential form
\begin{equation}
    \mathcal Q^{\mathcal{N}}(T)
    \simeq
    \exp\left[\mu(\overline{\mathcal M}-\mathcal I)\right]
    \circ\mathcal U(T)
\end{equation}
is recovered as the dilute-layer limit $q\to0$, $Nq=\mu$. Thus the Poissonian form is a useful approximation, while the binomial channel is the more general finite-depth expression.

The onset of the stationary-channel window occurs, in the worst case, at the scale
\begin{equation}
    T_{\rm stat}\sim \mu T^\ast=NqT^\ast,
\end{equation}
with high-probability finite-depth corrections of order $\sqrt{n_{O}/n_{\rm loc}}$. We emphasize that this is a worst-case estimate and that the noise model can become stationary for observables of interest at a much lower scale, with $T^{*}$ providing the characteristic single-fault reference scale.

The practical implication is that, beyond the crossover and for moderate noise levels, the dominant effect of noise can be mitigated by calibrating observable-dependent rescaling parameters. The same contrast factor can then be applied over an extended part of the dynamical curve. This provides a physical basis for calibration-based rescaling and for global-folding zero-noise extrapolation in Trotterized dynamics. The resulting simplified error mitigation procedure can provide up to an exponential advantage over conventional approaches in the regime of substantial noise. The observed crossover also controls the complexity of multiexponential ZNE, showing that error mitigation procedures can depend qualitatively on the overall unitary evolution. However, it also demonstrates that naive ZNE based on oversimplified functions can fail because increasing the noise level can drive the simulation across the crossover into the Zeno-like regime.

Fixed-depth simulations should be viewed as complementary to fixed-step and fixed-error simulations. Fixed-step and fixed-error protocols are useful for controlling discretization effects, minimizing unnecessary circuit depth, and validating product-formula convergence. Fixed-depth simulations are useful when one wants a time trace whose leading noise channel is stable across the endpoint-time scan. Combining these protocols can therefore give both accuracy checks and a practical route to noise mitigation.

Future work should explore how the memory time $T^\ast$ depends on integrability, chaos, and dimensionality, and how the rescaling picture extends beyond local depolarizing noise to more realistic device-level noise models. Another important direction is a fully local space-time formulation of the theory for large systems at fixed microscopic gate error probability, where the probability of at least one fault somewhere in a whole layer need not be small. We are also interested in obtaining tighter, model-dependent scaling laws for the time $T_{\rm stat}$.

\begin{acknowledgments}
We thank Artem Soloviev and Terentii Yaropolov for useful discussions.
\end{acknowledgments}

\section*{Data availability}
Data supporting the numerical figures in this paper are available from the corresponding author upon reasonable request. The code is available
at https://zenodo.org/records/21455333.

\bibliographystyle{apsrev4-2}
\bibliography{references_APS_PhysRevA_stylechecked}

\appendix

\section{Details of error operators and circuit-dependent effects}
\label{app:transpilation_effects}

This appendix records the circuit-level assumptions used in Eq.~\eqref{eq:fault_layer_factorization_new}. The main derivation does not require a particular four-qubit example of error propagation. It requires only the algebraic statement that a local Pauli fault can be moved to the beginning of a Trotter layer at the cost of a propagated Pauli string and a small coherent correction.

\subsection{Error propagation}

Consider a Trotter layer built from Clifford gates and Pauli rotations
\begin{equation}
    R_{\Pi_j}(\phi_j)=\exp(-i\phi_j\Pi_j),
    \qquad
    \phi_j=\omega_j\Delta,
    \label{eq:pauli_rotation_def}
\end{equation}
where $\Pi_j$ is a local Pauli string. A Pauli error $E$ is propagated through such gates by the elementary rules
\begin{equation}
    E R_{\Pi_j}(\phi_j)
    =
    R_{\Pi_j}(s_{E,j}\phi_j)E,
    \qquad
    s_{E,j}=
    \begin{cases}
        +1, & [E,\Pi_j]=0,\\
        -1, & \{E,\Pi_j\}=0,
    \end{cases}
    \label{eq:er_pass_rotation}
\end{equation}
and
\begin{equation}
    E C=C(C^\dagger E C)
    \label{eq:er_pass_clifford}
\end{equation}
for a Clifford gate $C$. Since Clifford conjugation maps Pauli strings to Pauli strings, the operator obtained after propagating all Pauli faults in a layer to the beginning of that layer is again a Pauli string, denoted $P_\xi$.

The rotations remaining after this propagation may have signs different from those in the clean Trotter layer. If $\widetilde U_{\Delta,\xi}$ denotes the same circuit skeleton with these modified signs, then
\begin{equation}
    U_{\Delta,\xi}=\widetilde U_{\Delta,\xi}P_\xi
    =U_\Delta V_{\Delta,\xi}P_\xi,
    \label{eq:UN_factorization}
\end{equation}
where
\begin{equation}
    V_{\Delta,\xi}=U_\Delta^\dagger\widetilde U_{\Delta,\xi}
    =I-i\Delta v_\xi+O(\Delta^2).
    \label{eq:V_first_order_app}
\end{equation}
To first order in $\Delta$, $v_\xi$ is the difference between the modified and unmodified local generators of the layer. In the transverse-field Ising example this means a sum of terms of the form $(\widetilde h_j-h_j)X_j$ and $(\widetilde J_j-J_j)Z_jZ_{j+1}$, with $\widetilde h_j,\widetilde J_j=\pm h_j,\pm J_j$. Higher-order BCH commutators are included in the $O(\Delta^2)$ remainder.

The locality assumption used in the Loschmidt-echo interpretation is not automatic for every possible transpilation. We assume that, for the elementary fault patterns carrying the dominant probability weight,
\begin{equation}
    |{\rm supp}\,P_\xi|=O(1),
    \qquad
    \|v_\xi\|\le v_{\max}=O(1)
    \label{eq:locality_assumption_P_v}
\end{equation}
independently of the number of Trotter layers. For CNOT layouts with long cascades, a local Pauli error can grow into a long string; such layouts fall outside the local single-fault version of the argument unless the corresponding growth is explicitly bounded. Native Pauli rotations such as $R_{ZZ}$ or $R_{ZX}$ usually make the locality of $P_\xi$ more transparent.

\subsection{Transpilation effects}
\label{app:transpilation}

The formal channel construction applies to any transpilation that satisfies the propagation and locality conditions above. The numerical value of the layer error probability $q$ and the distribution of propagated strings $P_\xi$, however, are circuit dependent.

For a circuit with angle-independent elementary error probabilities, the layer fault probability is
\begin{equation}
    q=1-\prod_{a\in {\rm layer}}(1-p_a)
    \simeq \sum_{a\in {\rm layer}}p_a
    \label{eq:q_general_layer_app}
\end{equation}
in the dilute-layer limit. If all elementary locations have the same error probability $p$, this reduces to Eq.~\eqref{eq:layer_error_probability_q}.

For native Pauli rotations, it is often useful to allow angle-dependent error probabilities. A simple phenomenological model is
\begin{equation}
    p_a(T)=p_{0,a}+\lambda_a |\omega_a|\frac{T}{N}.
    \label{eq:native_p_err}
\end{equation}
Then, in the dilute limit,
\begin{equation}
    q(T)\simeq q_0+\kappa\frac{T}{N},
    \qquad
    q_0=\sum_a p_{0,a},
    \qquad
    \kappa=\sum_a \lambda_a|\omega_a|.
    \label{eq:q_angle_dependent_app}
\end{equation}
The stationary-crossover condition becomes implicit,
\begin{equation}
    T_{\rm stat}\simeq N q(T_{\rm stat})T^\ast.
    \label{eq:Tstat_angle_implicit}
\end{equation}
Solving Eq.~\eqref{eq:Tstat_angle_implicit} gives
\begin{equation}
    T_{\rm stat}
    \simeq
    \frac{Nq_0T^\ast}{1-\kappa T^\ast},
    \label{eq:Tstar_renorm}
\end{equation}
provided $\kappa T^\ast<1$. The perturbative interpretation as a small renormalization of the layer error dose requires the stronger condition $\kappa T^\ast\ll1$. If the exact layer-fault
probability $q(T)=1-\prod_a[1-p_a(T)]$ is not small, Eq.~\eqref{eq:Tstat_angle_implicit} remains the appropriate finite-depth estimate, but the Poisson approximation should not be used.

\subsection{Arbitrary-order Trotter circuits}
\label{app:trot_ord}

For a product formula of order $g$, the coherent Trotter error scales as $T^{g+1}/N^g$, up to model-dependent constants. The channel derivation is unchanged in structure: one defines the faulty layer, propagates Pauli faults to the beginning of that layer, obtains $P_\xi$ and $V_{\Delta,\xi}$, constructs the single-fault channels $\mathcal M_\xi$, and then performs the binomial fixed-$M$ summation.

For example, for $H=H_A+H_B$ the second-order symmetric layer is
\begin{equation}
    U_\Delta^{(2)}
    =
    \exp\left(-i\frac{\Delta}{2}H_A\right)
    \exp\left(-i\Delta H_B\right)
    \exp\left(-i\frac{\Delta}{2}H_A\right).
    \label{eq:second_order_layer}
\end{equation}
There is no extra factor of $T$ in Eq.~\eqref{eq:second_order_layer}, because $\Delta=T/N$ already denotes the physical Trotter step.

The propagated strings $P_\xi$ and their probabilities need not be literally identical for different product-formula orders or transpilation layouts. What is invariant is the construction. If the dominant propagated strings remain local and the coherent corrections $V_{\Delta,\xi}$ satisfy the bounds of Appendix~\ref{app:simplify_single_error}, then the main result becomes
\begin{equation}
    \mathcal Q^{\mathcal N}(T)
    \simeq
    \left[(1-q^{(g)})\mathcal I+q^{(g)}\overline{\mathcal M}^{(g)}\right]^N
    \circ\mathcal U(T),
    \label{eq:final_channel_order_g}
\end{equation}
with the order- and layout-dependent layer error probability $q^{(g)}$ and averaged stationary channel $\overline{\mathcal M}^{(g)}$.

\section{Noise channels and error-summation technique}
\label{app:error_summation_technique}

This appendix gives the technical derivation behind Eq.~\eqref{eq:final_channel}. The proof is organized around two independent steps. First, the coherent intra-layer corrections $V_{\Delta,\xi}$ are shown to be perturbative. Second, for a fixed number of faulty layers, the ordered sum over their positions is replaced by stationary single-fault channels. The statistics of the number of faulty layers is then kept binomial, with the Poisson form obtained only as an additional limit.

\subsection{Neglecting coherent intra-layer corrections}
\label{app:simplify_single_error}

For a branch with $M$ nontrivial fault patterns, let $W_{\xi_1\ldots \xi_M}$ be the exact unitary obtained from the factorization
\begin{equation}
    U_{\Delta,\xi_j}=U_\Delta V_{\Delta,\xi_j}P_{\xi_j},
    \qquad
    V_{\Delta,\xi_j}=I-i\Delta v_{\xi_j}+O(\Delta^2),
    \label{eq:branch_exact_factor_app}
\end{equation}
and let $W^{(0)}_{\xi_1\ldots \xi_M}$ be the same branch with every $V_{\Delta,\xi_j}$ replaced by the identity. Since all surrounding factors are unitary, a telescoping estimate gives
\begin{equation}
    \left\|W_{\xi_1\ldots \xi_M}-W^{(0)}_{\xi_1\ldots \xi_M}\right\|
    \le
    \sum_{j=1}^M\left\|V_{\Delta,\xi_j}-I\right\|.
    \label{eq:telescoping_V_bound}
\end{equation}
If $\|v_{\xi_j}\|\le v_{\max}$, then
\begin{equation}
    \left\|W-W^{(0)}\right\|
    \lesssim
    M\Delta v_{\max}+O(M\Delta^2).
    \label{eq:V_branch_bound}
\end{equation}
For the corresponding unitary channels, this implies, up to an unimportant numerical factor,
\begin{equation}
    \left\|\mathcal W-\mathcal W^{(0)}\right\|_\diamond
    \lesssim
    M\frac{T}{N}v_{\max}+O\!\left(M\frac{T^2}{N^2}\right).
    \label{eq:V_channel_bound}
\end{equation}

For a single fault this correction is $O(Tv_{\max}/N)$ and vanishes in the continuous-step limit at fixed endpoint time. For a binomial ensemble, the typical number of faulty layers is $M\simeq Nq$, so the averaged scale of the neglected coherent correction is
\begin{equation}
    qT\bar v,
    \qquad
    \bar v=\sum_{\xi\ne0}w_\xi\|v_\xi\|.
    \label{eq:V_average_scale}
\end{equation}
Thus the Pauli-insertion channel derived in the main text is the leading term when
\begin{equation}
    qT\bar v\ll1.
    \label{eq:V_drop_condition}
\end{equation}
We note that first-order Trotter circuits that precede or lie within the stationary rescaling regime automatically satisfy \eqref{eq:V_drop_condition}. This follows directly from the conditions \eqref{eq:full_region_system_basic}: the maximum admissible $T_m$, which bounds the stationary rescaling region, is determined by the equation $T_{m} = \mu T^{*} = \alpha^{1/g+1}(\mu/q)^{g/g+1}$, which gives $T_{m} = \alpha / (qT^{*})^{g}$. For $g = 1$, we obtain $qT\bar v \lesssim qT_{m}\bar v \sim \alpha/T^{*} \ll 1$. The last strong inequality follows from the simple condition $\alpha \ll 1$, which is needed for small Trotterization error, where we consider nonvanishing $T^{*} \sim O(1)$. We also used that $\bar v$ is $O(1)$ since it is proportional to the $\sum_{n}J_{n}/N$, and we are using dimensionless units normalized by the same average coupling. For higher-order Trotter circuits, the validity of the condition \eqref{eq:V_drop_condition} should be checked explicitly. If necessary, it should be added to the stationary-region validity criteria as the condition $T < 1/q$, corresponding to a horizontal boundary line in diagrams of the type shown in Fig.~\ref{fig:rescaling_region}.

Even if condition ~\eqref{eq:V_drop_condition} is not satisfied, the formal structure of the derivation can still be retained by defining a $\Delta$-dependent single-fault channel built from $V_{\Delta,\xi}P_\xi$ rather than from $P_\xi$ alone. The resulting stationary channel would then be $\overline{\mathcal M}_\Delta$ instead of $\overline{\mathcal M}$, where $\Delta$ could be treated as a ``slow'' variable for the sufficiently deep circuits considered here.

\subsection{Single-fault channels: time invariance and convergence}
\label{app:channel_properties}

In this section, we prove the existence of the stationary single-fault channel limit $\mathcal{M}_{\xi}$ and establish a direct relation between its convergence rate, the Loschmidt echo, and the two-point-measurement work statistics of the $P_{\xi}$ quench. We also prove properties required for the isolation of the total noise channel in Eq.~\eqref{eq:final_channel}.

\subsubsection{Stationary channel existence and basis for memory loss}
We first note that, from a formal perspective, the stationary limit of $\mathcal{M}_{\xi}$ always exists and is given by the energy-dephased projection. This result follows directly by inserting a resolution of the identity into Eq.~\eqref{eq:M_xi_t_new}:

\begin{equation}
    \begin{gathered}
        \lim_{T\rightarrow\infty}\mathcal{M}_{\xi}^{(T)}(\hat{\rho}) = \lim_{T\rightarrow\infty}\sum_{m, k, l, n}\frac{1}{T}\int_{0}^{T} dt\langle m|P_{\xi}|k\rangle \langle l|P_{\xi}|n\rangle \exp(i(E_{m} - E_{k} + E_{l} - E_{n})t) |m\rangle \langle k|\hat{\rho}|l\rangle \langle n| = \\
        \sum_{E_{m} + E_{l} - E_{k} - E_{n} = 0}|m\rangle\langle n| \langle m|P_{\xi}|k\rangle \langle l|P_{\xi}|n\rangle\langle k|\hat{\rho}|l\rangle.
    \end{gathered}
    \label{eq:mean_T_channel}
\end{equation}
However, the convergence toward this limit can depend sensitively on the spectral properties of the system, and one may therefore be interested in a quantitative analysis of its rate. To quantify this convergence, one can use the normalized Frobenius superoperator norm of the mismatch channel $\mathcal{M}_{\xi}^{(T)}-\mathcal{M}_{\xi}$ \cite{Nielsen2021gatesettomography}. This is a proper norm on the linear space of quantum maps and provides bounds on observables and probabilities. For an arbitrary channel $\Lambda$, a normalized Frobenius norm can be defined via the corresponding Pauli-basis transfer matrix:

\begin{equation}
    \begin{gathered}
    ||\Lambda||_{2}^{2} = \frac{1}{d^2}\sum_{k,l}\Lambda^{\dagger}_{kl}\Lambda_{lk} = \frac{1}{d^2}\sum_{k,l}\Lambda^{*}_{lk}\Lambda_{lk},\\
    \Lambda_{lk} = \frac{\mathrm{Tr}\left(\sigma_{l}\Lambda\left(\sigma_{k}\right)\right)}{d}.
    \end{gathered}
    \label{eq:channel_frobenius_def}
\end{equation}

For left-right maps $\Phi_{A,B}$, $\Phi_{A,B}(\hat{\rho}) = \Phi_{c}A\hat{\rho}B$, where $\Phi_{c}$ is a complex number, the definition \eqref{eq:channel_frobenius_def} gives a closed expression for the introduced norm in terms of the $\mathrm{Tr}(C^{\dagger}D)$ scalar product on the space of operators:

\begin{equation}
    \begin{gathered}
    ||\Phi_{A,B}||_{2}^{2} =  \frac{1}{d^2}\Phi_{c}\Phi_{c}^{*}\mathrm{Tr}\left(AA^{\dagger}\right)\mathrm{Tr}\left(BB^{\dagger}\right).
    \end{gathered}
    \label{eq:left-right_norm}
\end{equation}
On the linear space of left-right maps, one can define the corresponding Hilbert-Schmidt product, which, as follows from \eqref{eq:left-right_norm}, induces the Frobenius norm discussed above:

\begin{equation}
    \begin{gathered}
    \langle \Phi_{A,B}|\Phi_{C, D}\rangle_{HS} = \frac{1}{d^2}\Phi^{*}_{A,B}\Phi_{C,D}\mathrm{Tr}(A^{\dagger}C)\mathrm{Tr}(B^{\dagger}D), \\
    ||\Phi - \Xi||_{2}^{2} =  \langle \Phi - \Xi|\Phi - \Xi\rangle_{HS}.
    \end{gathered}
    \label{eq:channel_norm_definition}
\end{equation}

Defining standard ``jump operators'' $P_{\omega} = \sum_{E_{n} - E_{k} = \omega}\langle n|P_{\xi}|k\rangle|n\rangle\langle k|$, we get representations of the channels $\mathcal{M}_{\xi}^{(T)}$ and $\mathcal{M}_{\xi}$ in terms of left-right maps:

\begin{equation}
    \mathcal{M}_{\xi}^{(T)}(\hat{\rho}) = \frac{1}{T}\int_{0}^{T}dt\sum_{\omega, \nu}e^{i(\omega + \nu)t}P_{\omega}\hat{\rho}P_{\nu},
    \label{eq:left-right_repr}
\end{equation}

\begin{equation}
    \mathcal{M}_{\xi}(\hat{\rho}) = \sum_{\omega + \nu = 0}P_{\omega}\hat{\rho}P_{\nu}.
    \label{eq:stationary_jump}
\end{equation}

Using \eqref{eq:left-right_repr}, \eqref{eq:stationary_jump}, and the sesquilinearity of the Hilbert-Schmidt product \eqref{eq:channel_norm_definition} on the space of left-right maps, one obtains a spectral representation of the mismatch norm $||\mathcal{M}_{\xi}^{(T)} - \mathcal{M}_{\xi}||_{2}^{2} = \Delta_{\mathcal{M}}(T)$:

\begin{equation}
    \Delta_{\mathcal{M}}(T) = \frac{1}{d^{2}T^2}\int_{0}^{T}\int_{0}^{T}dt'dt''\sum_{\omega, \nu}e^{i(\omega + \nu)\left(t' - t''\right)}\mathrm{Tr}(P^{\dagger}_{\omega}P_{\omega})\mathrm{Tr}(P^{\dagger}_{\nu}P_{\nu}) - \frac{1}{d^2}\sum_{\omega + \nu = 0}\mathrm{Tr}(P^{\dagger}_{\omega}P_{\omega})\mathrm{Tr}(P^{\dagger}_{\nu}P_{\nu}),
    \label{eq:mismatch_norm_start}
\end{equation}
where we have used the orthogonality relation $\mathrm{Tr}(P_{\omega}^{\dagger}P_{\nu})\propto \delta_{\omega\nu}$. To obtain a convenient continuous representation of the mismatch norm in Eq.~\eqref{eq:mismatch_norm_start}, which later allows us to take the thermodynamic limit and work with a continuous spectrum, we introduce the distribution

\begin{equation}
    z(\Omega) = \frac{1}{d}\sum_{\omega}\delta(\Omega - \omega)\mathrm{Tr}(P^{\dagger}_{\omega}P_{\omega}).
    \label{eq:work_distribution_def}
\end{equation}

Since $P_{\xi}$ is unitary, it is easy to see that $z(\Omega)$ is a proper probability distribution: $\int z(\Omega)d\Omega = \mathrm{Tr}(\sum_{\omega}P^{\dagger}_{\omega}P_{\omega})\frac{1}{d} = \mathrm{Tr}(I / d) = 1$. Using the fact that $P_{\xi}$ is also Hermitian, which implies the property $P^{\dagger}_{\omega} = P_{-\omega}$, one can also show that $z(\Omega)$ is always symmetric around zero:
\begin{equation}
    z(-\Omega) = \frac{1}{d}\sum_{\omega}\delta(\Omega + \omega)\mathrm{Tr}(P^{\dagger}_{\omega}P_{\omega}) = \frac{1}{d}\sum_{-\omega}\delta(\Omega - (-\omega))\mathrm{Tr}(P^{\dagger}_{-\omega}P_{-\omega}) = z(\Omega).
\end{equation}

Replacing the spectral sums in \eqref{eq:mismatch_norm_start} by integration over frequencies, we obtain

\begin{equation}
    \Delta_{\mathcal{M}}(T) = \frac{1}{T^2}\int_{0}^{T}\int_{0}^{T}dt'dt''\int_{-\infty}^{\infty}e^{i\Omega(t' - t'')}z\ast z(\Omega)d\Omega - \lim_{\epsilon\rightarrow 0}\int_{-\epsilon}^{\epsilon}z\ast z(\Omega)d\Omega,
    \label{eq:mismatch_core_expression}
\end{equation}
where $z*z(\Omega)$ denotes the standard convolution convention:
\begin{equation}
    z*z(\Omega) = \int z(\omega)z(\Omega - \omega)d\omega.
\end{equation}

We are now ready to obtain the final energy representation of the mismatch norm. Integrating over time in \eqref{eq:mismatch_core_expression}, we obtain

\begin{equation}
    \Delta_{\mathcal{M}}(T) = \int \frac{4\sin^2\left(\frac{\Omega T}{2}\right)}{\Omega^2 T^2}\overline{z}_{\rm res}(\Omega)d\Omega.
    \label{eq:energy_repr}
\end{equation}
Here, the probability distribution $\overline{z}(\Omega)=z\ast z(\Omega)$ describes the sum of two independent random variables distributed according to $z(\Omega)$. The subscript ${\rm res}$ indicates that a possible singular contribution at $\Omega=0$ has been removed, i.e., that the corresponding Dirac-delta peak is subtracted whenever it is present. The convergence of the channel $\mathcal{M}_{\xi}^{(T)}$ can then be studied in the energy representation using Eq.~\eqref{eq:energy_repr}. At large times, the asymptotic behavior of $\Delta_{\mathcal{M}}(T)$ is determined by the behavior of the residual distribution $\overline{z}_{\rm res}(\Omega)$ near $\Omega=0$, and therefore by the thermodynamic-limit form of $z(\Omega)$.

To proceed, we need to clarify the physical meaning of $z(\Omega)$ and derive a time-domain representation of Eq.~\eqref{eq:mismatch_core_expression}. Integrating over the energy variable first, we obtain

\begin{equation}
    \Delta_{\mathcal{M}}(T) = \frac{1}{T^2}\int_{0}^{T}\int_{0}^{T}dt'dt''|\chi(t' - t'')|^2 - \lim_{\epsilon\rightarrow 0}\int_{-\epsilon}^{\epsilon}z\ast z(\Omega)d\Omega,
    \label{eq:time_repr_start}
\end{equation}
where $\chi(t)$ is the characteristic function of the $z(\Omega)$ distribution. Here we use the standard property that the characteristic function of a symmetric distribution is purely real. Using the definition \eqref{eq:work_distribution_def}, we arrive at

\begin{equation}
    \begin{gathered}
    \chi(t) = \mathrm{Tr}\left(\sum_{\omega}e^{i\omega t}P^{\dagger}_{\omega}P_{\omega}\right)\frac{1}{d} = \mathrm{Tr}\left(\sum_{\omega}P_{-\omega}\exp(iHt)P_{\omega}\exp(-iHt)\right)\frac{1}{d} = \\ \mathrm{Tr}\left(\left(\sum_{\omega}P_{-\omega}\right)\exp(iHt)\left(\sum_{\nu}P_{\nu}\right)\exp(-iHt)\frac{I}{d}\right) = \\
    \mathrm{Tr}\left(P_{\xi}\exp(iHt)P_{\xi}\exp(-iHt)\frac{\exp\left(-\beta H\right)}{Z_{\beta}}\right)_{\beta=0} = \left\langle \exp(i(H + \delta E)t)\exp(-iHt)\right\rangle_{\beta = 0} = G(t),
    \end{gathered}
    \label{eq:LE_char}
\end{equation}
where we have used the same orthogonality property $\mathrm{Tr}(P^{\dagger}_{\omega}P_{\nu}) = \mathrm{Tr}(P_{-\omega}P_{\nu}) \sim \delta_{\omega\nu}$ in the transition from the first to the second line in \eqref{eq:LE_char} and $\delta E=P_{\xi}HP_{\xi}-H$ is a local perturbation that depends on the error pattern $\xi$. The relation $\chi(t)=G(t)$, derived in Eq.~\eqref{eq:LE_char}, shows that the characteristic function of the distribution $z(\Omega)$ coincides with the infinite-temperature Loschmidt echo generated by the $P_{\xi}$ error quench. In the theory of quantum work statistics, the Loschmidt echo is identified with the characteristic function of the corresponding work distribution~\cite{Work_LE}. Thus, $z(\Omega)$ can be interpreted as the infinite-temperature two-point-measurement work distribution. Calculating and analyzing the $z(\Omega)$ distribution is a broad topic in quantum thermodynamics. Relevant results for integrable quadratic and general many-body systems, for specified and random
quenches, can be found in \cite{tierz2025workstatisticssuddenquantum, _obejko_2017, PhysRevE.100.052136, PhysRevE.98.012106}.

Let us make several general statements about the mismatch norm using the energy representation \eqref{eq:energy_repr}. If $\overline{z}_{\rm res}(\Omega)$ is analytic near zero, the asymptotic form of $\Delta_{\mathcal{M}}(T)$ is simply $\Delta_{\mathcal{M}}(T) \approx 2\pi\overline{z}_{\rm res}(0)/T$. Therefore, the large-$T$ regime is fixed by analyticity alone. However, we are interested in estimating $T^{*}_{\xi}$, which corresponds to the crossover time scale at which $\Delta_{\mathcal{M}}(T)$ enters the aforementioned regime. For this purpose, one can use a Gaussian approximation to obtain a qualitatively correct dependence of the mismatch norm. This is possible because all moments $\varsigma_{z}^{m}$ of the distribution $z(\Omega)$ can be calculated via commutator expansions:

\begin{equation}
    \varsigma_{z}^{m} = \int z(\Omega)\Omega^{m}d\Omega = \frac{1}{d}\sum_{\omega}\omega^{m}\mathrm{Tr}(P^{\dagger}_{\omega}P_{\omega}) = \frac{1}{d}\sum_{\omega}\mathrm{Tr}(P^{\dagger}_{\omega}\mathrm{Ad}^{m}_{H}(P_{\omega})) = \frac{1}{d}\mathrm{Tr}(P_{\xi}\mathrm{Ad}^{m}_{H}(P_{\xi})),
    \label{eq:z_moment_formula}
\end{equation}
which for the case $m = 2$ reduces to:
\begin{equation}
    \varsigma_{z}^{2} = \sigma_{z}^{2} = \frac{1}{d}\mathrm{Tr}([P_{\xi}, H][H, P_{\xi}]).
\end{equation}
For typical errors in the CNOT-transpilation of the TFIM model, $\sigma_{z} \sim J_{n}, h_{n}$. Since $\overline{z}(\Omega)$ and $z(\Omega)$ in this approximation are both Gaussian with moment relation $\sigma_{\overline{z}} = \sqrt{2}\sigma_{z}$, after inserting the distribution into the formula \eqref{eq:energy_repr} we obtain:

\begin{equation}
\Delta_{\mathcal M}(T)
\sim
\frac{\sqrt{\pi}}{\sigma_{z} T}
\operatorname{erf}\!\left(\sigma_{z}T\right)
-
\frac{1}{\sigma_{z}^2 T^2}
\left(
1-e^{-\sigma_{z}^2 T^2}
\right).
\label{eq:mismatch_gaussian}
\end{equation}
We emphasize that \eqref{eq:mismatch_gaussian} and other moment-based distribution approximations do not describe the asymptotic behavior of $\Delta_{\mathcal{M}}(T)$ as $T\rightarrow \infty$. The large-$T$ regime is determined only by the analytic form of the distribution $\overline{z}(\Omega)$ near zero, whereas moments capture only integral properties. Nevertheless, such approximations can correctly indicate the crossover scale in $T$, since they remain valid for small and intermediate values of $T$. Thus, using the qualitative approximation \eqref{eq:mismatch_gaussian}, one can identify the corresponding time scale $T_{\xi}^{*}$, which controls the crossover into the asymptotic $1/T$ regime, as $T_{\xi}^{*} \sim 1/\sigma_{z}$.

If the thermodynamic limit is nonstandard and yields a distribution $z(\Omega)$ that is not square-integrable, $\overline{z}_{\rm res}(\Omega)$ can be nonanalytic near zero. In this case, the asymptotic form of $\Delta_{\mathcal{M}}(T)$ is again determined by the type of divergence or nonregularity at this point. Since $\overline{z}_{\rm res}(\Omega)$ is a proper distribution, we state that a typical nonanalyticity near zero can be represented in the general form $\overline{z}_{\rm res}(\Omega) \sim |\Omega|^{\beta}|\ln(|\Omega|)|^{\gamma}$ with $-1 < \beta < 1$ and $\gamma > 0$, thereby accounting for possible van Hove-type singularities. For such singularities, we obtain the asymptotic behavior

\begin{equation}
    \Delta_{M}(T) \sim |\ln(T)|^{\gamma}T^{-\beta - 1}.
\end{equation}
Thus, in most cases, we have at worst effectively algebraic decay of the mismatch norm.

Now, we shift our discussion back to the time-domain
representation of the mismatch norm. Applying the same analysis as in \eqref{eq:LE_char} to the singular term in Eq.~\eqref{eq:time_repr_start} and changing variables in the integral, we obtain:

\begin{equation}
    \begin{gathered}
    \Delta_{\mathcal{M}}(T) = \frac{2}{T^2}\int_{0}^{T}(T - t)\left(|G(t)|^{2} - |\mathcal{G}|^2\right)dt = 2\int_{0}^{1}(1 - x)(|G(xT)|^2 -|\mathcal{G}|^2)dx, \\
    G(t) = \left\langle \exp(i(H + \delta E)t)\exp(-iHt)\right\rangle_{\beta = 0},
    \end{gathered}
    \label{eq:mismatch_LE}
\end{equation}
where $|\mathcal{G}|^2$ denotes the stationary plateau of the time-averaged Loschmidt echo signal:

\begin{equation}
    \begin{gathered}
    |\mathcal{G}|^2 = \lim_{T\rightarrow\infty}\frac{1}{T}\int_{0}^{T}G(t)G^{*}(t)dt = \frac{1}{d^2}\sum_{w = \nu}\mathrm{Tr}(P^{\dagger}_{\omega}P_{\omega})\mathrm{Tr}(P^{\dagger}_{\nu}P_{\nu}) = \frac{1}{d^2}\sum_{\omega = \nu}\mathrm{Tr}(P^{\dagger}_{\omega}P_{\omega})\mathrm{Tr}(P^{\dagger}_{-\nu}P_{-\nu}) = \\
    \frac{1}{d^2}\sum_{\omega + \nu = 0}\mathrm{Tr}(P^{\dagger}_{\omega}P_{\omega})\mathrm{Tr}(P^{\dagger}_{\nu}P_{\nu}),
    \end{gathered}
    \label{eq:LE_plato}
\end{equation}
where we have used the cyclicity of the trace, the property of the jump operators $P_{\nu}^{\dagger}=P_{-\nu}$, and a change of variables in the term $\mathrm{Tr}(P_{\nu}^{\dagger}P_{\nu})$. Equation~\eqref{eq:mismatch_LE} establishes a connection between the relaxation of the Loschmidt echo and the convergence of the single-error channel.

The Loschmidt echo signal generated by a local quench in a many-spin system with local interactions can exhibit different large-time regimes depending on ergodicity, integrability, disorder, and the structure of the many-body spectrum. In weakly perturbed ergodic systems, the decay may follow a Fermi-golden-rule-type exponential form, while strongly interacting many-spin systems can also display perturbation-independent decay controlled by the intrinsic interaction time scale rather than by a semiclassical escape-rate mechanism~\cite{Gorin_2006_LE, LE_report, Zangara2016, Sanchez2020, Fine2014}. Critical or integrable one-dimensional local quenches may instead exhibit algebraic or logarithmic asymptotics, together with light-cone-related finite-size revival structures~\cite{StephanDubail2011}. Highly excited spin chains may nevertheless show a transient exponential local-echo decay instead of the low-energy orthogonality-catastrophe power law~\cite{LupoSchiro2016}. In many-body-localized systems, the relevant long-time behavior is slower: Loschmidt-echo fluctuations can decay algebraically, while related spin-echo observables can saturate~\cite{SerbynAbanin2017}. Finite-size systems additionally exhibit saturation and revival effects; therefore, finite-size echoes should be interpreted with care when extracting thermodynamic asymptotics~\cite{Zangara2016}. Thus, we do not assume a universal asymptotic form of the Loschmidt echo relaxation, treating the large-time approach to the stationary contribution as a model-dependent dephasing property.

Equation~\eqref{eq:mismatch_LE} thus reduces the convergence problem for the single-fault channel to the long-time relaxation properties of the corresponding local Loschmidt echo, providing a theoretical basis for estimating $T^{*}_{\xi}$ from the Loschmidt-echo crossover. For a broad class of regimes of interest, we argue that the effective asymptotic behavior of the mismatch norm is at worst $\Delta_{\mathcal{M}}(T) \sim T^{\beta}$ with $\beta < 0$, yielding qualitative convergence of $\mathcal{M}_{\xi}^{(T)}$ to the stationary channel $\mathcal{M}_{\xi}$.

\subsubsection{Stationary channel time-invariance}

Now we show that $\mathcal{M}_{\xi}$ commutes with the full unitary evolution channel $\mathcal{U}$, which is equivalent to invariance of the channel under the unitary dynamics of the system, i.e., under time translation. Using a change of variables, we obtain

\begin{widetext}

\begin{equation}
    \label{eq:time_inv_proof}
    \begin{gathered}
        \mathcal{M}_{\xi} \circ \mathcal{U}_{T'}(\hat{\rho}) = \lim_{T\rightarrow\infty}\frac{1}{T}\int_{0}^{T}dt\exp\left(-iHt\right)P_{\xi}\exp\left(iHt\right)\exp\left(-iHT'\right) \hat{\rho}
        \exp\left(iHT'\right)\exp\left(-iHt\right)P_{\xi}\exp\left(iHt\right) = \\ \lim_{T\rightarrow\infty}\frac{1}{T}\exp\left(-iHT'\right)
        \int_{0}^{T}dt\exp\left(-iH(t - T')\right)P_{\xi}\exp\left(iH(t - T')\right)\hat{\rho}
        \exp\left(iH(T' - t)\right)P_{\xi}\exp\left(iH(t - T')\right)\exp\left(iHT'\right) = \\ \mathcal{U}_{T'} \circ \mathcal{M}_{\xi}(\hat{\rho}),
    \end{gathered}
    \end{equation}
\end{widetext}
where existence of the limit is validated by the previous discussion on convergence of $\mathcal{M}_{\xi}$ operators. For the commutator of these channels, we can explicitly write:

\begin{equation}
    ||\mathcal{M}^{(T)}_{\xi} \circ \mathcal{U}(T) - \mathcal{U}(T) \circ \mathcal{M}^{(T)}_{\xi}||_{2} \leq 2||(\mathcal{M}^{(T)}_{\xi} - \mathcal{M}_{\xi})||_{2}  = 2\sqrt{\Delta_{\mathcal{M}}(T)},
\end{equation}
where we used Frobenius norm invariance under unitary conjugation. Thus, the accuracy of the time-invariance statement \eqref{eq:time_translation_invariance_main} is also controlled by the Loschmidt echo relaxation properties.

This time-invariance is important for the many-error summation procedures, defined in the next section.

\subsubsection{Edge effects in
single-fault channel averaging}

Finally, it is important to qualitatively understand the origin and conditions of fault positions averaging into a stationary channel $\mathcal{M}_{\xi}$ which we extensively use further. Since any error can happen at any layer of the considered circuit, there are always errors occurring very close to the endpoint time $T$, such that $|T - T_{\rm err}| < T^{*}$. Even though we explicitly account for all error positions in the original definition of one-fault contributions $\mathcal{S}_{\xi}^{(T)}$ \eqref{eq:one_err_mean_pre_stationary}, intuitively, one can expect such edge-faults to seriously influence the overall noise channel. We examine this issue below.

Let us consider the normalized one-fault contribution which accounts for faults occurring in the time window $(S_{1}(T), S_{2}(T))$, where $S_{1,2}(T) = s_{1,2}T,$ and $0< s_{1} < s_{2} < 1$ are fractions that define omitted edges. After the continuum limit, this results in a simple formula for the final one-fault channel:

\begin{equation}
    \begin{gathered}
    \mathcal{\widetilde{S}}_{\xi}^{(T)} = Nr_{\xi}(1 - q)^{N - 1}\mathcal{U}(T)\circ \mathcal{\widetilde{M}}_{\xi}^{(T)}, \\
    \mathcal{\widetilde{M}}_{\xi}^{(T)} = \mathcal{U}(-S_{1})\circ\mathcal{M}_{\xi}^{(S_{2} - S_{1})}\circ\mathcal{U}(S_{1}). 
    \end{gathered}
    \label{eq:edge_omitted_one_fault} 
\end{equation}

Applying the same analysis as for the mismatch norm of the full channel  $\mathcal{M}_{\xi}^{(T)}$, one can obtain:

\begin{equation}
    ||\mathcal{\widetilde{M}}_{\xi}^{(T)} - \mathcal{M}_{\xi}||^{2}_{2} = \Delta_{\mathcal{M}}\left((s_{2} - s_{1})T\right).
    \label{eq:edge_omitted_norm}
\end{equation}

The result for the partial averaged fault contribution mismatch norm \eqref{eq:edge_omitted_norm} sheds light on the importance of edge
effects. In the practically relevant case of $s_{2} - s_{1} \sim 1$, which is correct for $T \gg T^{*}$, using the time representation of $\Delta_{\mathcal{M}}(T)$ \eqref{eq:mismatch_LE}, it is easy to estimate that difference between full and partial one-error channels mismatch norms is controlled by the Loschmidt echo relaxation properties at the endpoint:

\begin{equation}
    \big|||\mathcal{\widetilde{M}}_{\xi}^{(T)} - \mathcal{M}_{\xi}||^{2}_{2} - ||\mathcal{M}_{\xi}^{(T)} - \mathcal{M}_{\xi}||^{2}_{2}\big| \sim (1 - (s_{2} - s_{1}))^{2}\times |\langle G^{2}(T) - |\mathcal{G}|^{2}\rangle|,
    \label{eq:mismatch_difference}
\end{equation}
where $\langle\dots \rangle$ stands for the small-oscillation averaged (envelope) quantity. Since norm difference in \eqref{eq:mismatch_difference}, in most cases, is at least exponentially suppressed, we can conclude that omitting the edge faults does not significantly affect the total noise channel. Using the partial average $\mathcal{\widetilde{M}_{\xi}}^{(T)}$ instead of the full average $\mathcal{M}_{\xi}^{(T)}$ introduces a negligible change in the overall accuracy, and thus, they are equivalent in terms of used norm.

The same reasoning applies to the ideal evolution channel commutator norm:

\begin{equation}
    ||\mathcal{\widetilde{M}}^{(T)}_{\xi} \circ \mathcal{U}(T) - \mathcal{U}(T) \circ \mathcal{\widetilde{M}}^{(T)}_{\xi}||_{2} \leq 2||(\mathcal{\widetilde{M}}^{(T)}_{\xi} - \mathcal{M}_{\xi})||_{2}  = 2\sqrt{\Delta_{\mathcal{M}}((s_{2} - s_{1})T)}.
\end{equation}

Thus, we can conclude that in the $T \gg T^{*}$ regime only "bulk" averaging truly influences the stationary channel limit.

\subsection{\texorpdfstring{Fixed-$M$ ordered sums}{Fixed-M ordered sums}}
\label{app:error_summ}

We derive here the stationary contribution of a fixed number $M$ of faulty layers.
This step is independent of the subsequent choice of the probability distribution
of $M$. The latter may be binomial, Poissonian, or any other distribution
supported on the number of faulty layers. Here we focus on the conservative noncommuting propagated-error case only.

Let $\xi_1,\ldots,\xi_M$ be nontrivial fault patterns
at fixed positions within their Trotter layers, ordered from the
beginning to the end of the circuit. For a given ordered set of faulty layers
$1\le m_1<\cdots<m_M\le N$, and after neglecting the intra-layer coherent corrections $V_{\Delta,\xi}$ discussed above, the corresponding branch channel can be
written in the toggling frame as
\begin{equation}
    \mathcal W_{m_1\ldots m_M}^{(0)}
    =
    \mathcal P_{\xi_M}(t_{m_M})
    \circ\cdots\circ
    \mathcal P_{\xi_1}(t_{m_1})
    \circ \mathcal U(T),
    \qquad
    t_m=m\Delta ,
    \label{eq:branch_channel_toggling_app}
\end{equation}
where $\Delta=T/N$ and
\begin{equation}
    \mathcal P_\xi(t)(\hat{\rho})
    =
    P_\xi(-t)\hat{\rho} P_\xi(-t)
    \label{eq:toggling_insertion_channel_app}
\end{equation}
with the sign convention used in the single-fault channel. In Eq.~\eqref{eq:branch_channel_toggling_app}, we have already used the time-invariance property of the $\mathcal{M}_{\xi}$ channels in Eq.~\eqref{eq:time_inv_proof}, which holds after the positional memory-loss time $T^{*}$. This transformation is valid only in the context of the error-summation procedure described below.

The contribution of this fixed ordered pattern, before summing over different $\xi_{m}$ pattern combinations, is
\begin{equation}
    \mathcal S_{\xi_1\ldots\xi_M}^{(N)}
    =
    (1-q)^{N-M}\prod_{j=1}^M r_{\xi_j}
    \sum_{1\le m_1<\cdots<m_M\le N}
    \mathcal W_{m_1\ldots m_M}^{(0)} .
    \label{eq:fixed_pattern_discrete_sum_app}
\end{equation}
Here $r_\xi$ is the probability of the nontrivial layer pattern $\xi$, so that
$q=\sum_{\xi\ne0}r_\xi$ and $w_\xi=r_\xi/q$.

The mechanism behind the averaging is the following. Once an error is moved to
the toggling frame, its microscopic layer position is converted into the time
argument of a dressed local operator $P_\xi(-t)$. The random layer position
therefore samples the orbit of this local operator under the ideal Hamiltonian.
For times longer than the memory time $T_{\xi}^{*}$, the Cesaro average of the corresponding
insertion channel approaches the stationary single-fault channel:
\begin{equation}
    \mathcal M_{\xi}
    =
    \lim_{T\to\infty}
    \frac{1}{T}\int_0^T dt\,\mathcal P_\xi(t).
    \label{eq:stationary_single_fault_app}
\end{equation}

Conditioned on having exactly $M$ faulty layers, their positions are uniformly
distributed over all ordered subsets of $\{1,\ldots,N\}$. Hence the normalized
discrete sum is an average over order statistics. In the continuum limit,
\begin{equation}
    \frac{1}{\binom{N}{M}}
    \sum_{m_1<\cdots<m_M}
    \mathcal W_{m_1\ldots m_M}^{(0)}
    \longrightarrow
    \frac{M!}{T^M}
    \int_{0<t_1<\cdots<t_M<T}
    dt_1\cdots dt_M\,
    \mathcal W_{t_1\ldots t_M}^{(0)} .
    \label{eq:ordered_simplex_app}
\end{equation}
The ordered simplex is the only remnant of the time ordering of the faults.

We now show why this ordered average factorizes into stationary single-fault
channels. Define
\begin{equation}
    I_M(T)
    =
   \int_{0<t_1<\cdots<t_M<T}
    dt_1\cdots dt_M\,
    \mathcal P_{\xi_M}(t_M)
    \circ\cdots\circ
    \mathcal P_{\xi_1}(t_1).
    \label{eq:IM_definition_app}
\end{equation}
For $M=1$, once $T > T^{*}$ and the $\mathcal{M}_{\langle\xi\rangle}^{(T)}$ channels have crossed into the stationary regime, we obtain
\begin{equation}
    I_1(T)
    =
    T\mathcal M_{\xi_1}
    +
    o(T).
    \label{eq:I1_app}
\end{equation}
For larger $M$ one may use the recursive identity
\begin{equation}
    I_M(T)
    =
    \int_0^T dt\,
    \mathcal P_{\xi_M}(t)
    \circ I_{M-1}(t).
    \label{eq:IM_recursion_app}
\end{equation}
Assuming
\begin{equation}
    I_{M-1}(t)
    =
    \frac{t^{M-1}}{(M-1)!}
    \mathcal M_{\xi_{M-1}}\circ\cdots\circ\mathcal M_{\xi_1}
    +
    o(t^{M-1}),
    \label{eq:IM_induction_assumption_app}
\end{equation}
Using stationarity,
\begin{equation}
    \frac{M}{T^M}
    \int_0^T dt\, t^{M-1}\mathcal P_{\xi_M}(t)
    =
    \mathcal M_{\xi_M}
    +
    o(1),
    \label{eq:weighted_cesaro_app}
\end{equation}
one obtains
\begin{equation}
    I_M(T)
    =
    \frac{T^M}{M!}
    \mathcal M_{\xi_M}\circ\cdots\circ\mathcal M_{\xi_1}
    +
    o(T^M).
    \label{eq:IM_stationary_app}
\end{equation}
Thus, for fixed $M$,
\begin{equation}
    \frac{M!}{T^M} I_M(T)
    \simeq
    \mathcal M_{\xi_M}\circ\cdots\circ\mathcal M_{\xi_1}.
    \label{eq:ordered_simplex_factorization_app}
\end{equation}

Equivalently,
\begin{equation}
    \frac{1}{\binom{N}{M}}
    \sum_{m_1<\cdots<m_M}
    \mathcal W_{m_1\ldots m_M}^{(0)}
    \simeq
    \mathcal M_{\xi_M}\circ\cdots\circ\mathcal M_{\xi_1}
    \circ\mathcal U(T).
    \label{eq:fixed_pattern_stationary_app}
\end{equation}
The approximation requires that the ordered positions have enough time to
sample the stationary orbit of each insertion. In the notation of the main
text this condition is
\begin{equation}
    \frac{T}{M}\gtrsim T^{*},
    \label{eq:fixed_M_self_average_condition_app}
\end{equation}
up to a tolerance-dependent numerical factor. The exceptional regions of the
simplex in which one of the insertions lies too close to a boundary, or in
which two insertions are separated by a time shorter than the memory time, have
small relative volume in this regime. The replacement of sums by integrals gives an additional
finite-step correction of order $O_M(1/N)$, or equivalently $O(M/N)$ at the
level of the crude scaling estimate.

Substituting Eq.~\eqref{eq:fixed_pattern_stationary_app} into
Eq.~\eqref{eq:fixed_pattern_discrete_sum_app} gives the fixed-pattern
binomial contribution
\begin{equation}
    \mathcal S_{\xi_1\ldots\xi_M}^{\rm bin}(T)
    \simeq
    \binom{N}{M}
    (1-q)^{N-M}
    \prod_{j=1}^M r_{\xi_j}
    \,
    \mathcal M_{\xi_M}\circ\cdots\circ\mathcal M_{\xi_1}
    \circ\mathcal U(T).
    \label{eq:fixed_pattern_binomial_app}
\end{equation}
Using $r_\xi=qw_\xi$ and summing over all nontrivial patterns yields
\begin{equation}
    \mathcal S_M^{\rm bin}(T)
    \simeq
    \binom{N}{M}q^M(1-q)^{N-M}
    \left(\sum_{\xi\ne0}w_\xi\mathcal M_\xi\right)^M
    \circ\mathcal U(T).
    \label{eq:SM_binomial_app}
\end{equation}
This is the fixed-$M$ sector used in the main text. Summing
Eq.~\eqref{eq:SM_binomial_app} over $M$ gives the stationary binomial global
channel,
\begin{equation}
    \mathcal Q^{\mathcal N}(T)
    \simeq
    \left[
    (1-q)\mathcal I
    +
    q\overline{\mathcal M}
    \right]^N
    \circ\mathcal U(T),
    \qquad
    \overline{\mathcal M}
    =
    \sum_{\xi\ne0}w_\xi\mathcal M_\xi.
    \label{eq:binomial_global_channel_app}
\end{equation}
The Poissonian expression follows from this result in the dilute-layer limit
$q\to0$, $Nq=\mu$,
\begin{equation}
    \left[
    (1-q)\mathcal I
    +
    q\overline{\mathcal M}
    \right]^N
    =
    \left[
    \mathcal I
    +
    q(\overline{\mathcal M}-\mathcal I)
    \right]^N
    \longrightarrow
    \exp\!\left[\mu(\overline{\mathcal M}-\mathcal I)\right].
    \label{eq:poisson_from_binomial_app}
\end{equation}

\subsection{Binomial distribution, Poisson limit, and applicability criteria}
\label{app:poisson_lim}

In this section we estimate which values of $M$ dominate the binomial distribution and hence determine the stationary crossover scale.

Let $M$ be a binomial random variable with parameters $(N,q)$. A Bernstein--Chernoff bound gives
\begin{equation}
    {\rm Pr}\{M-Nq\ge x\}
    \le
    \exp\left[-\frac{x^2}{2Nq(1-q)+2x/3}\right].
    \label{eq:chernoff_bernstein_app}
\end{equation}
Therefore, with probability at least $1-\delta$,
\begin{equation}
    M\le M_\delta
    =
    Nq+
    O\!\left(\sqrt{Nq(1-q)\ln(1/\delta)}+\ln(1/\delta)\right).
    \label{eq:M_delta_app}
\end{equation}
The corresponding high-probability stationary onset is
\begin{equation}
    T_{\rm stat}(\delta)
    \sim
    T^\ast\left[Nq+
    O\!\left(\sqrt{Nq(1-q)\ln(1/\delta)}+\ln(1/\delta)\right)\right].
    \label{eq:Tstat_highprob_main}
\end{equation}
At leading order this reduces to $T_{\rm stat}\sim NqT^\ast$. If one accounts for the required absolute accuracy $\delta \sim a_{O}(N,q) \sim (1 - p)^{n_{O}N}$, this results in the correction $T_{\rm stat}\sim Nq\left(1 + O\left(\sqrt{\frac{n_{O}}{n_{\rm loc}}}\right)\right)T^\ast$.

The Poisson channel follows from Eq.~\eqref{eq:binomial_global_channel_app} only when $q\to0$ with $\mu=Nq$ fixed. Writing $A=\overline{\mathcal M}-\mathcal I$, one has
\begin{equation}
    \left[\mathcal I+qA\right]^N
    =
    \exp\left[
    \mu A-\frac{\mu q}{2}A^2+O(\mu q^2\|A\|^3)
    \right]
    \label{eq:poisson_correction_app}
\end{equation}
whenever $q\|A\|<1$. Thus the Poisson expression is accurate on a given sector when the correction $\mu q\|A\|^2$ is small, or more weakly on observable modes for which $A$ has small effective eigenvalue. For fixed $q$ and $N\to\infty$, the correct stationary channel is instead the binomial channel of Eq.~\eqref{eq:final_channel}.

The ordered-sum continuum approximation imposes an additional requirement. Typical sectors have $M\simeq Nq$, so the finite-step error in Eq.~\eqref{eq:fixed_pattern_stationary_app} is of order $M/N\simeq q$. Hence the layer-level proof is controlled in the dilute-layer regime $q\ll1$, even though the binomial expression itself is the correct finite-$q$ summation once the fixed-$M$ factorization is assumed. If the physical limit is taken at fixed local gate error probability $p$ while the number of locations per layer grows without bound, then $q=1-(1-p)^{n_{\rm loc}}$ approaches unity. In that thermodynamic setting the natural variables are local fault densities in space-time, not the probability that an entire layer contains at least one fault.

Combining the stationary onset with the coherent product-formula accuracy gives the necessary window
\begin{equation}
    T\gtrsim T_{\rm stat}(\delta),
    \qquad
    \frac{T^{g+1}}{N^g}<\alpha.
    \label{eq:full_applicability_window_app}
\end{equation}
The first condition is a conservative sufficient requirement for self-averaging, and the second is the usual Trotter accuracy requirement.

\subsection{Error-channel spectral properties and observable-level depolarization}
\label{app:M_final_prop}

Here we give a short spectral bound that justifies the observable-level depolarization approximation used in the main text. Let us use the previously introduced normalized Frobenius superoperator norm
\begin{equation}
    \|\Lambda\|_{2}^{2}
    =
    \frac{1}{d^{2}}\operatorname{Tr}
    \left(\Lambda^{\dagger}\Lambda\right),
    \label{eq:frob_norm_global_app}
\end{equation}
where trace is taken with respect to the Pauli superoperator basis. This norm is equivalent to Eq.~\eqref{eq:channel_frobenius_def} written in an orthonormal Hilbert--Schmidt operator basis. The stationary one-fault channel has the Bohr-frequency representation
\begin{equation}
    \overline{\mathcal M}(\hat\rho)
    =
    \sum_{\xi\ne0}w_{\xi}
    \sum_{\omega}P_{\xi,\omega}\hat\rho P_{\xi,\omega}^{\dagger}.
    \label{eq:Mbar_bohr_global_app}
\end{equation}
It is a self-adjoint contraction on operator space. We therefore write
\begin{equation}
    \overline{\mathcal M}(O_j)=\lambda_j O_j,
    \qquad
    -1\leq\lambda_j\leq1,
    \qquad
    \lambda_1=1,
    \label{eq:Mbar_spectrum_global_app}
\end{equation}
where $O_1=I/\sqrt d$ is the identity mode, and all $O_{j}$ are chosen to be Hermitian.

Microscopic fault labels that give the same propagated Pauli ray must first be grouped. Identifying $P$ and $-P$, define
\begin{equation}
    \widetilde w_P
    =
    \sum_{\xi:\,P_{\xi}=\pm P}w_{\xi},
    \qquad
    W_2=\sum_P\widetilde w_P^{2}.
    \label{eq:grouped_weight_global_app}
\end{equation}
The first moment of $\overline{\mathcal M}$ is zero, which follows directly from the 1-design properties of the Pauli basis. Thus, the second spectral moment is
\begin{equation}
    m_2(\overline{\mathcal M})
    =
    \frac{1}{d^{2}}\operatorname{Tr}
    \left(\overline{\mathcal M}^{2}\right)
    =
    \frac{1}{d^{2}}
    \sum_{\omega}\sum_{P,Q}
    \widetilde w_P\widetilde w_Q
    \left|\operatorname{Tr}
    \left(P_{\omega}^{\dagger}Q_{\omega}\right)\right|^{2}.
    \label{eq:m2_exact_global_app}
\end{equation}

To prove a useful inequality for the $\mathcal{\overline{M}}$ spectrum, we transfer to superoperator ket-bra notation. In this form, one can define weighted fault-ensemble operator $R$:
\begin{equation}
    |f_P\rangle=\frac{|P\rangle\rangle}{\sqrt d},
    \qquad
    R=\sum_P\widetilde w_P|f_P\rangle\langle f_P|,
    \qquad
    \mathcal P(R)=\sum_{\omega}\Pi_{\omega}R\Pi_{\omega},
    \label{eq:R_pinching_global_app}
\end{equation}
where $\Pi_{\omega}$ projects onto the operator subspace with Bohr frequency $\omega$:
\begin{equation}
    \Pi_{\omega}|f_{P}\rangle = \frac{|P_{\omega}\rangle\rangle}{\sqrt{d}}.
\end{equation}

Directly from Eq.~\eqref{eq:m2_exact_global_app},
\begin{equation}
    m_2(\overline{\mathcal M})
    =
    \|\mathcal P(R)\|_{\rm HS}^{2}.
    \label{eq:m2_pinched_global_app}
\end{equation}
Inserting the identity $\sum_{\omega} \Pi_{\omega} = \mathbf{1}$, we obtain the inequality:
\begin{equation}
    m_2(\overline{\mathcal M})
    =
    \|\mathcal P(R)\|_{\rm HS}^{2}
    \leq
    \|R\|_{\rm HS}^{2}
    =
    \sum_P\widetilde w_P^{2} = W_{2},
    \label{eq:pinching_bound_global_app}
\end{equation}
where in the dilute, low-error-per-location regime considered here, one can approximate $W_{2}$ via initial $P_{\xi}$ operators:

\begin{equation}
    W_{2} = \sum_{P}\widetilde{w}^{2}_{P} \approx \sum_{\xi, \zeta}w_{\xi}w_{\zeta}|\mathrm{Tr}(P_{\xi}P_{\zeta})|^{2}/d^{2}
\end{equation}
We assume that, for typical Hamiltonians and transpilations, there is no mechanism that concentrates the elements of the $R$ operator solely in the Bohr-diagonal sector. Thus, the inequality in Eq.~\eqref{eq:pinching_bound_global_app} can be parametrically strong and is often qualitatively closer to a $\ll$ bound.

For the Poisson channel, the eigenvalue on $O_j$ is
\begin{equation}
    a_j=e^{-\mu(1-\lambda_j)}.
    \label{eq:poisson_eigenvalue_global_app}
\end{equation}
The reference depolarizing contrast is $e^{-\mu}$. For $-1\leq\eta\leq1$,
\begin{equation}
    \left|e^{-\mu(1-\eta)}-e^{-\mu}\right|
    \leq
    \mu|\eta|,
    \label{eq:poisson_lipschitz_global_app}
\end{equation}
which follows from the mean-value theorem. Finally, we obtain:
\begin{align}
    \left\|\mathcal E_{\mu}^{\rm Pois}
    -\mathcal D_{e^{-\mu}}\right\|_{2}^{2}
    &\leq
    \frac{\mu^{2}}{d^{2}}\sum_{j}\lambda^{2}_{j} = \mu^{2}m_{2}(\overline{\mathcal{M}}).
    \label{eq:poisson_distance_global_app}
\end{align}
Thus the sufficient spectral-average condition for global depolarization is $\mu^{2}W_2\ll1$, or, in the plausible case, a weaker $\mu^{2}W_{2} \lesssim 1$. In the one-error-per-location limit, $W_2\sim n_{\rm loc}^{-1}$, giving the estimate used in Eq.~\eqref{eq:global_depol_req}.

The finite-depth binomial channel has eigenvalues
\begin{equation}
    b_j=\left[(1-q)+q\lambda_j\right]^N.
    \label{eq:binomial_eigenvalue_global_app}
\end{equation}
Its natural depolarizing reference is therefore $\mathcal D_{(1-q)^N}$. Since
\begin{equation}
    \left|\frac{d}{d\lambda}
    \left[(1-q)+q\lambda\right]^N\right|
    \leq Nq = \mu,
    \label{eq:binomial_lipschitz_global_app}
\end{equation}
we eventually obtain the bound of the same order for the binomial form of the stationary noise channel in \eqref{eq:final_channel}.

The same second-moment bound also gives a simple concentration statement, obtained from a Markov-type inequality:
\begin{equation}
    \frac{|\{j:\,|\lambda_j|\geq\epsilon\}|}{d^{2}}
    \leq
    \frac{W_2}{\epsilon^{2}}.
    \label{eq:spectral_tail_global_app}
\end{equation}
Thus, apart from a fraction at most $W_2/\epsilon^2$ of operator modes, the Poisson-channel eigenvalues lie between $e^{-\mu(1+\epsilon)}$ and $e^{-\mu(1-\epsilon)}$.

Finally, for an expansion $O=\sum_j o_jO_j$,
\begin{equation}
    \mathcal E_{N,q}^{\rm bin}(O)
    =\sum_j o_j\left[(1-q)+q\lambda_j\right]^N O_j,
    \qquad
    \mathcal E_{\mu}^{\rm Pois}(O)
    =\sum_j o_j e^{-\mu(1-\lambda_j)}O_j.
    \label{eq:modal_noise_global_app}
\end{equation}
Strong noise therefore selects the modes with the largest damping factors. The identity mode always survives, while additional slow modes are not excluded. In the large-depth limit $N\rightarrow\infty$, for traceless observables we obtain:

\begin{equation}
    \mathcal{E}_{N,q}^{\rm bin}(O) \approx (1 - q + q\lambda_{2})^{N}\sum_{n}o_{\lambda_{2n}}O_{\lambda_{2n}},
\end{equation}
where we account for the fact that the second-largest eigenvalue can be degenerate. Thus, in the dilute-layer limit considered here, the projection onto the $\lambda_{2}$ eigenspace controls the large-error behavior of observables in a Trotter circuit.

\newpage

\section{Numerical experiments across different physical regimes}
\label{app:numerical_sim}

In this final section, we extend our numerical analysis of noisy, fixed-depth Trotter circuits to a broader class of local Hamiltonians and physical regimes. Specifically, we investigate observable dynamics in the 1-D transverse-field and mixed-field Ising models (TFIM and MFIM), as well as in an XX transport model, exploring different coupling strengths and both integrable and non-integrable regimes. These additional simulations aim to demonstrate the robustness and generality of the stationary affine-rescaling behavior identified in the main text, showing that it persists beyond the specific models and parameter regimes considered previously.

We first examine the mixed-field Ising model (MFIM), obtained by extending the TFIM with an additional longitudinal field:
\begin{equation}
    H_{\mathrm{MFIM}} =
    J\sum_{j}Z_{j}Z_{j+1}
    + h\sum_{j}X_{j}
    + h_{z}\sum_{j}Z_{j},
    \label{eq: mixed-field_hamiltonian}
\end{equation}
where setting $h_z=0$ recovers the integrable TFIM, while a finite longitudinal field generically breaks integrability. To simulate this model, we employ the same first-order, fixed-depth Trotter decomposition introduced in Eq.~\eqref{eq:first_order_layer_AB}, supplemented by an additional layer of virtual $R_Z$ rotations implementing the longitudinal-field contribution.

Figures~\ref{fig:MFIM_J=4h_big_sim} and \ref{fig:MFIM_J=0.25h_big_sim} present the resulting observable dynamics for two distinct coupling regimes: deep in the interaction-dominated regime ($J=4h$) and deep in the field-dominated regime ($J=0.25h$), which correspond to ferromagnetic and paramagnetic physics. By additionally considering nonzero longitudinal fields, we probe the robustness of the observed behavior against integrability breaking. Remarkably, the qualitative stationary affine-rescaling effect remains clearly identifiable across all investigated regimes, suggesting that its emergence does not rely on the integrability of the underlying Hamiltonian.

We further investigate an XX transport model, supplemented by a next-nearest-neighbor $ZZ$ interaction that allows us to explicitly probe the effects of integrability breaking:
\begin{equation}
    H_T =
    J\sum_{j}(X_{j}X_{j+1} + Y_{j}Y_{j+1})
    + V_I\sum_{j}Z_{j}Z_{j+2}.
\end{equation}
The first term describes nearest-neighbor excitation transport, while the second introduces an effective density-density interaction in the fermionic representation, generically breaking the integrability of the non-interacting transport model.

The corresponding Trotter circuit retains the same general structure as in the preceding simulations. In particular, the nearest-neighbor $XX+YY$ evolution operators are compiled using suitable single-qubit rotations and only two CNOT gates, whereas the additional $Z_jZ_{j+2}$ interactions are implemented through a decomposition analogous to that employed for the Ising-type $ZZ$ couplings.

Figure~\ref{fig:XX+YY_big_sim} illustrates the resulting observable dynamics in noisy, fixed-depth Trotter circuits for this transport model. Once again, we observe qualitatively similar affine-rescaling behavior, consistent with the findings obtained for the Ising-type Hamiltonians.

Taken together, these numerical results provide further evidence that the stationary affine-rescaling phenomenon is not restricted to a particular Hamiltonian, coupling regime, or integrability class. Instead, it appears to be a robust feature of the noisy fixed-depth Trotter dynamics investigated in this work.

\newpage
\newcommand{\panelabelnumqubit}[1]{%
    \raisebox{0.0em}[0pt][0pt]{\textbf{(#1)}}\\[1.0em]
}

\begin{figure*}[h]
    \centering

    \begin{minipage}[t]{0.48\textwidth}
        \centering

        \includegraphics[width=\linewidth]{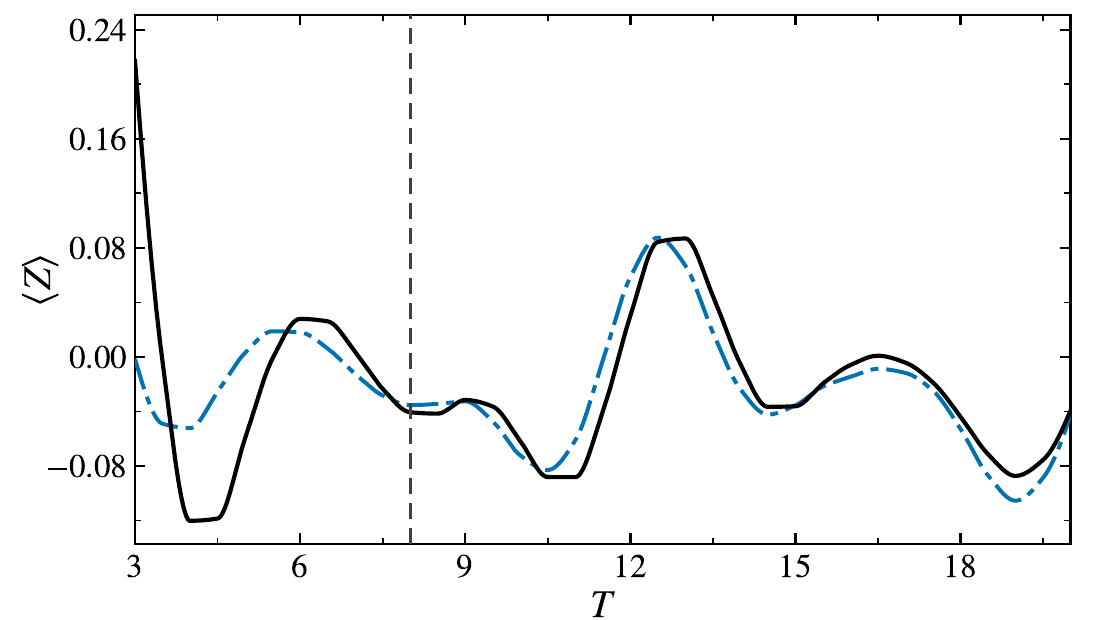}

        \includegraphics[width=\linewidth]{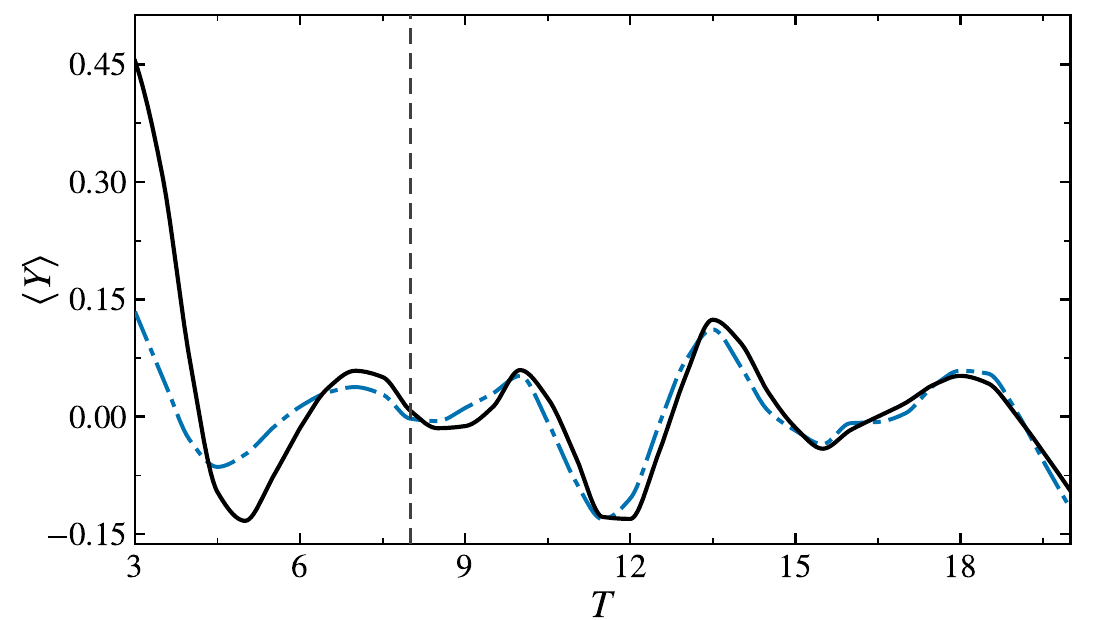}
        \panelabelnumqubit{$N_{q} = 8$}
        
        \includegraphics[width=\linewidth]{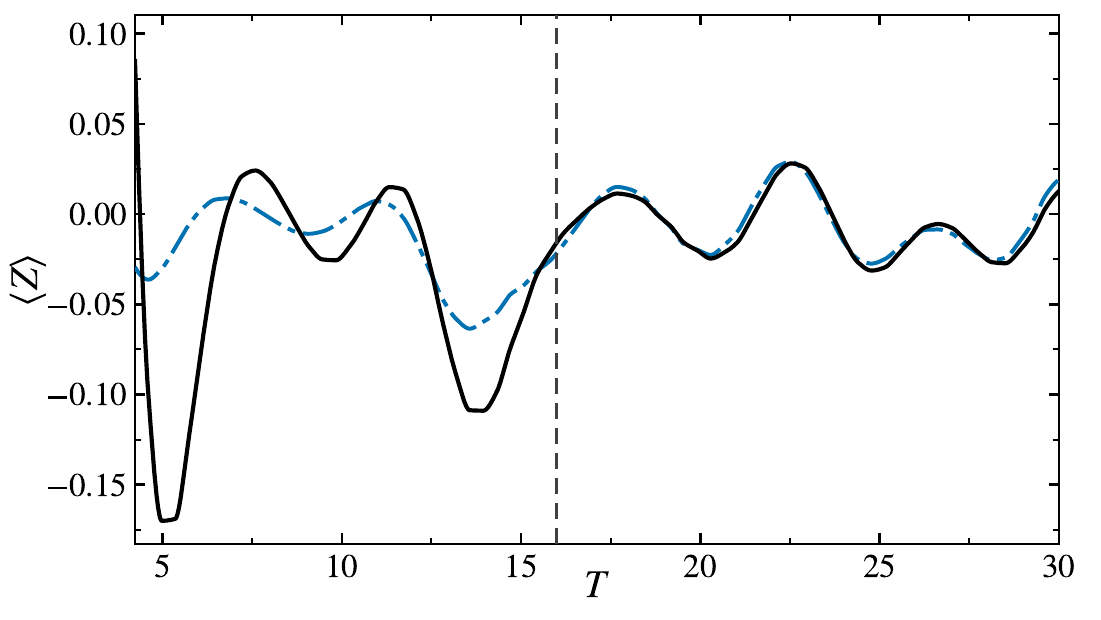}

        \includegraphics[width=\linewidth]{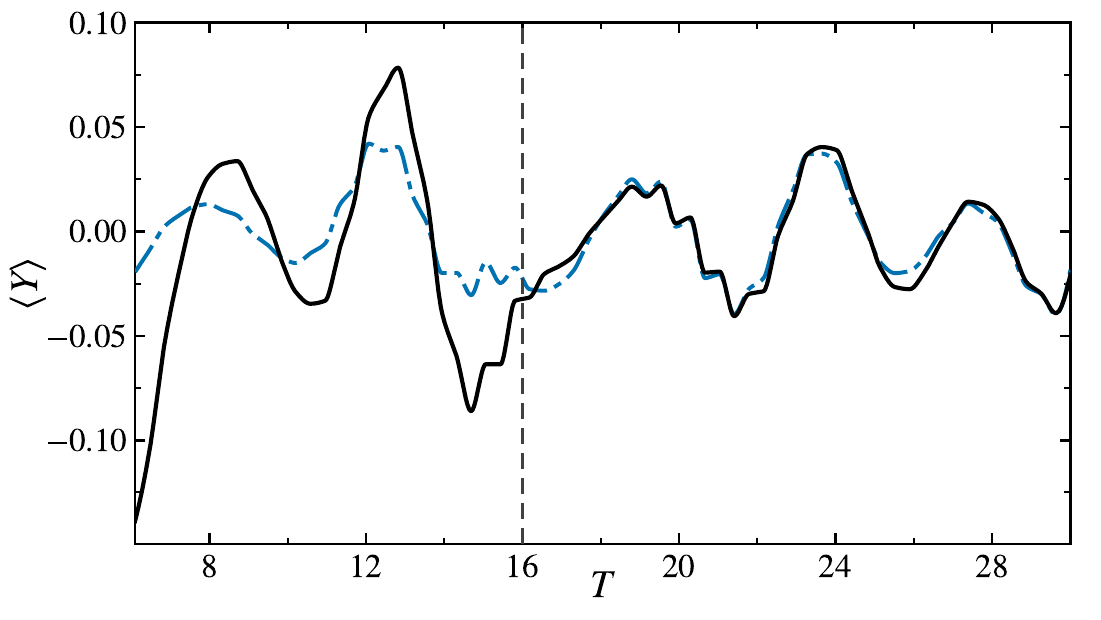}
        \panelabelnumqubit{$N_{q} = 10$}
    \end{minipage}
    \begin{minipage}[t]{0.48\textwidth}
        \centering

        \includegraphics[width=\linewidth]{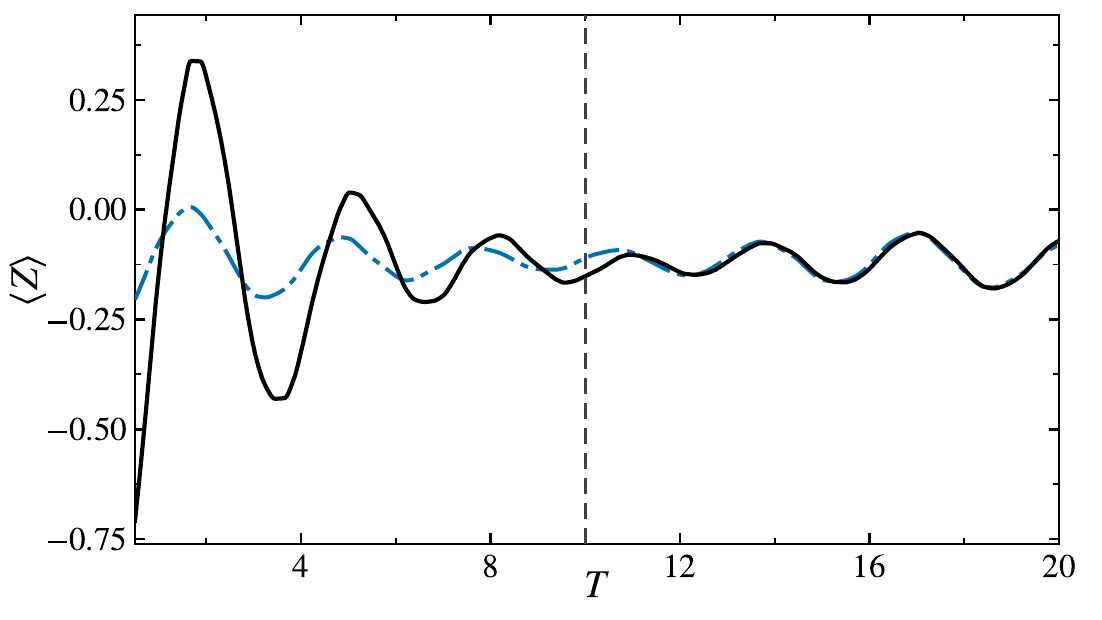}

        \includegraphics[width=\linewidth]{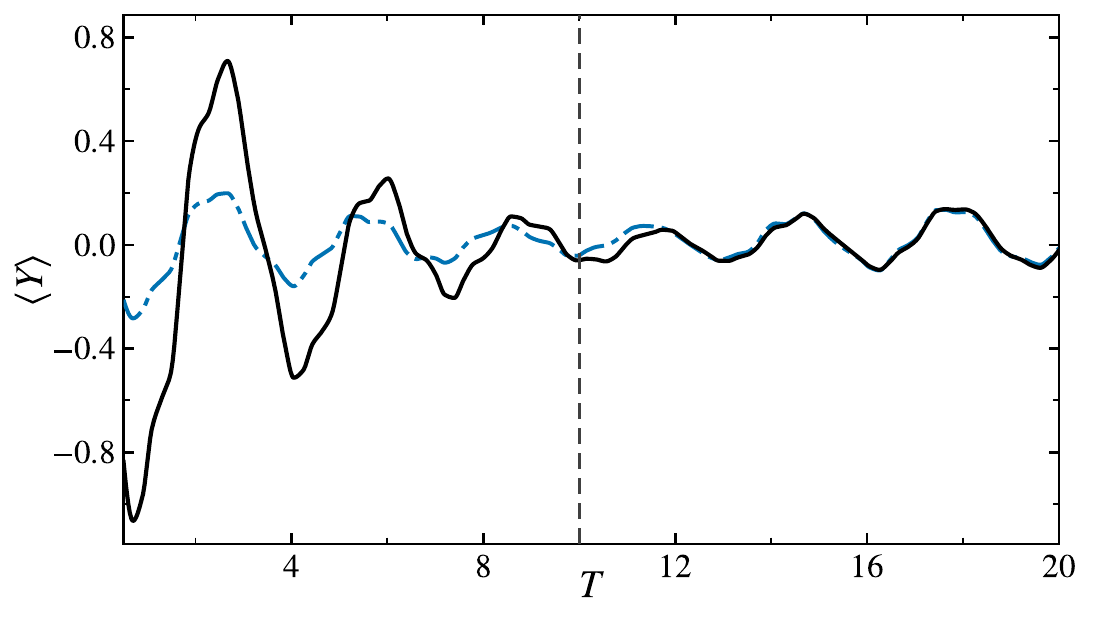}

        \panelabelnumqubit{$N_{q} = 8$}

        \includegraphics[width=\linewidth]{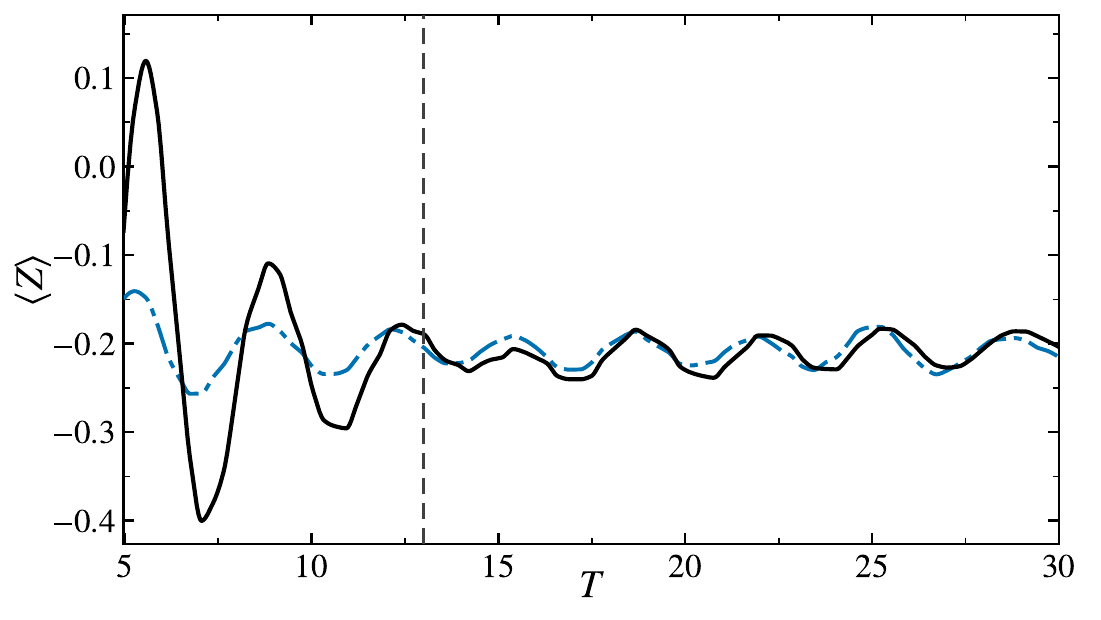}

        \includegraphics[width=\linewidth]{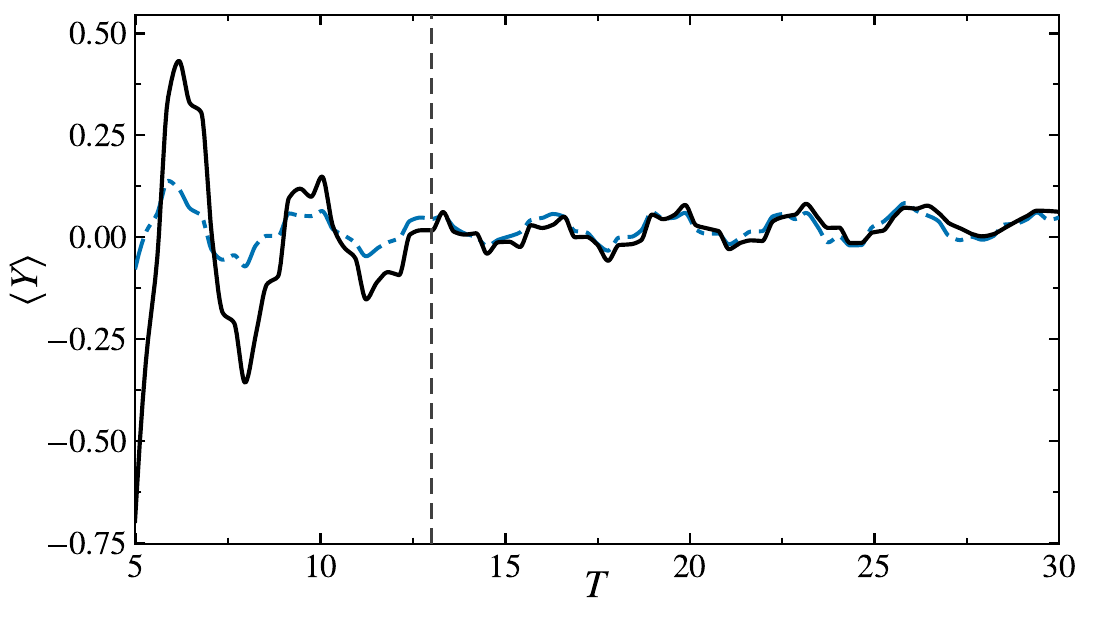}
        \panelabelnumqubit{$N_{q} = 10$}
    \end{minipage}
    \caption{The dynamics of the average magnetizations $\langle Z\rangle$ and $\langle Y\rangle$ in the MFIM model as a function of dimensionless time in the strongly ferromagnetic regime $J = 4h$ for 8- and 10- qubit number in fixed-depth Trotter circuits. The left column shows the integrable case $h_{z} = 0$, whereas the right column explores the non-integrable regime, which does
not admit a quadratic-fermion description, $|h_{z}| \sim J$. The initial states are $|10110110\rangle$ and $|1011011011\rangle$ respectively, the chain is homogeneous with $J_i=J$, $h_i=h$ and $h_{z,i} = h_{z}$. The dashed line shows the ideal Trotterized evolution for $N=100$ layers. The solid line is the affinely reconstructed noisy signal at gate error probability $p=0.001$ obtained via the same Trotter circuit.}
    \label{fig:MFIM_J=4h_big_sim}
\end{figure*}

\newpage
\begin{figure*}[h]
    \centering

    \begin{minipage}[t]{0.48\textwidth}
        \centering

        \includegraphics[width=\linewidth]{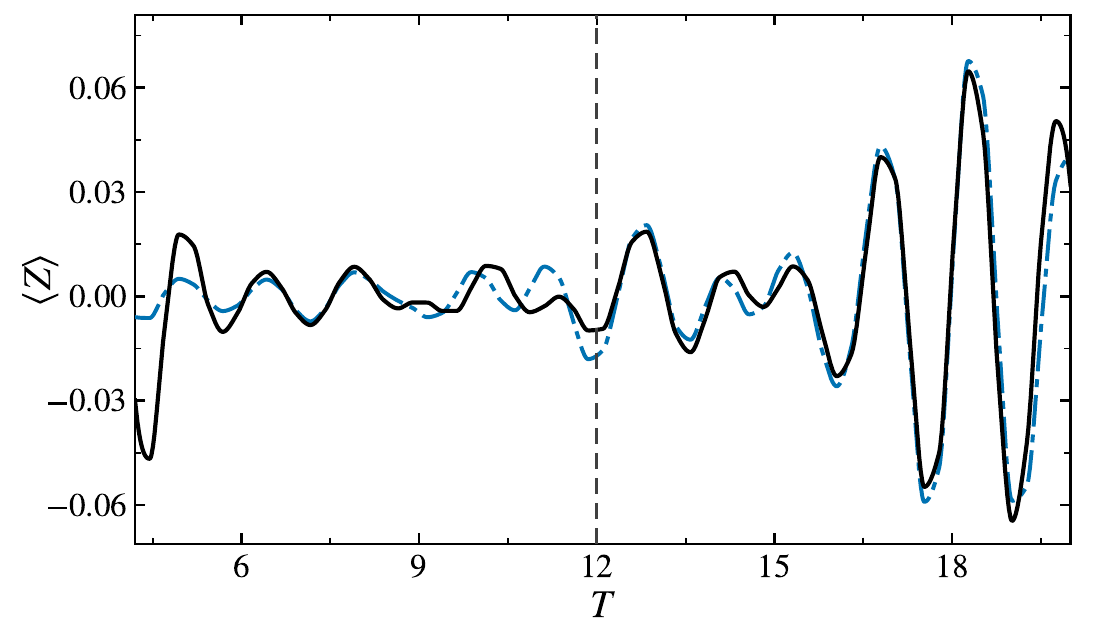}

        \includegraphics[width=\linewidth]{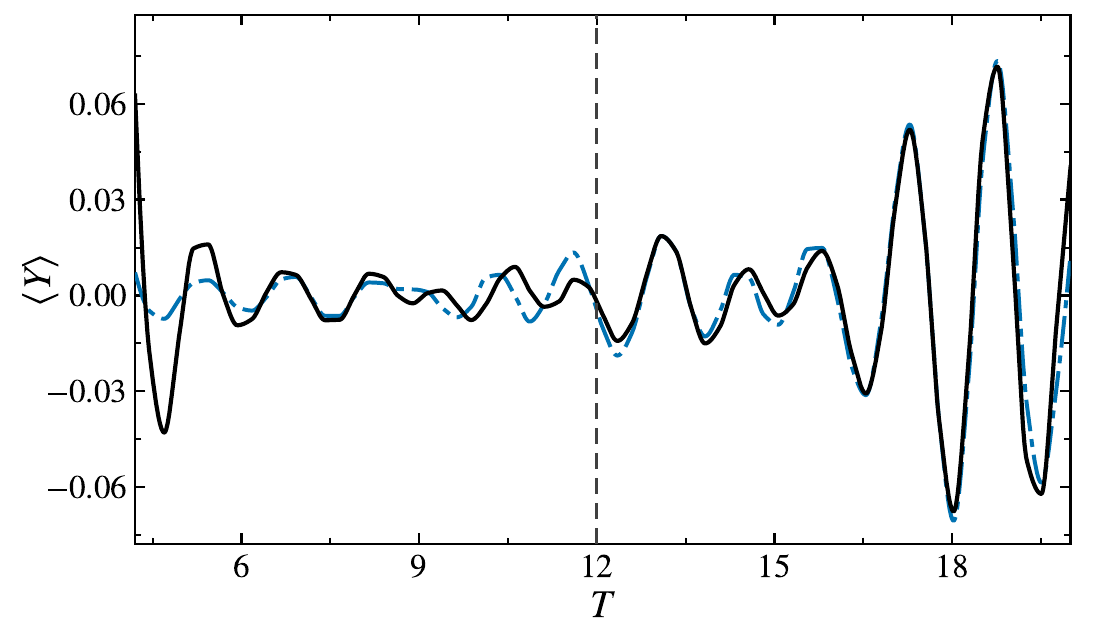}
        \panelabelnumqubit{$N_{q} = 8$}
        \includegraphics[width=\linewidth]{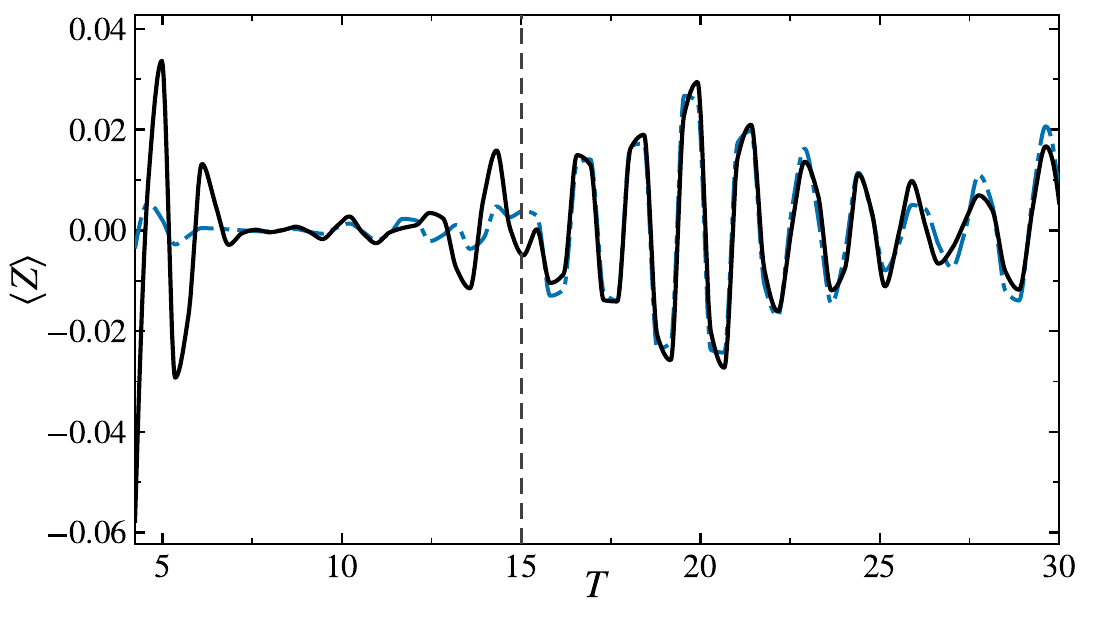}

        \includegraphics[width=\linewidth]{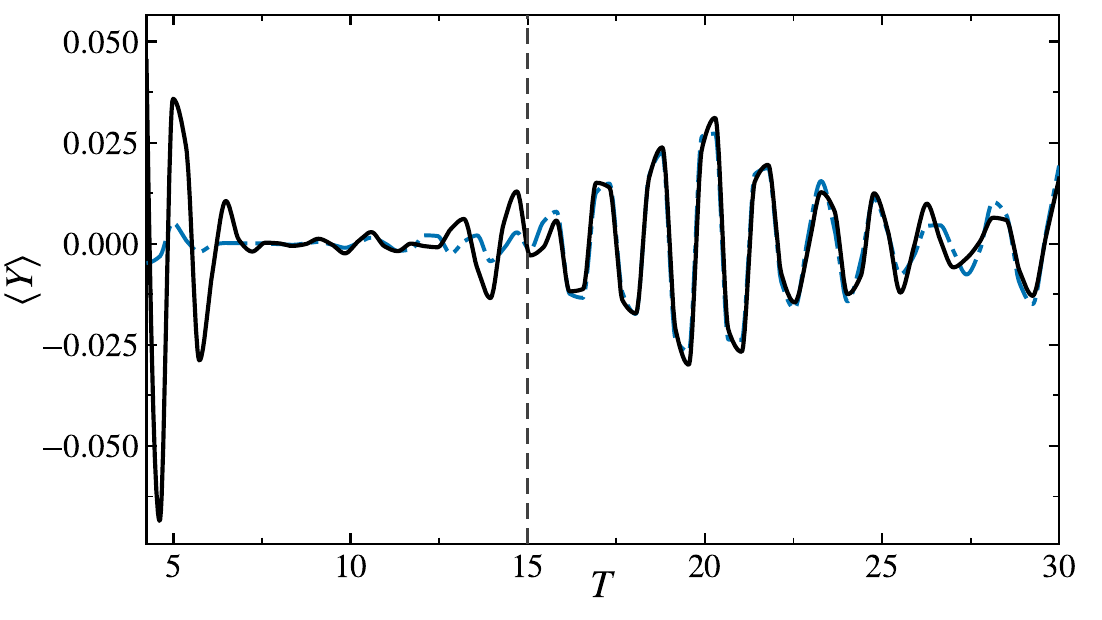}
        \panelabelnumqubit{$N_{q} = 10$}
    \end{minipage}
    \begin{minipage}[t]{0.48\textwidth}
        \centering

        \includegraphics[width=\linewidth]{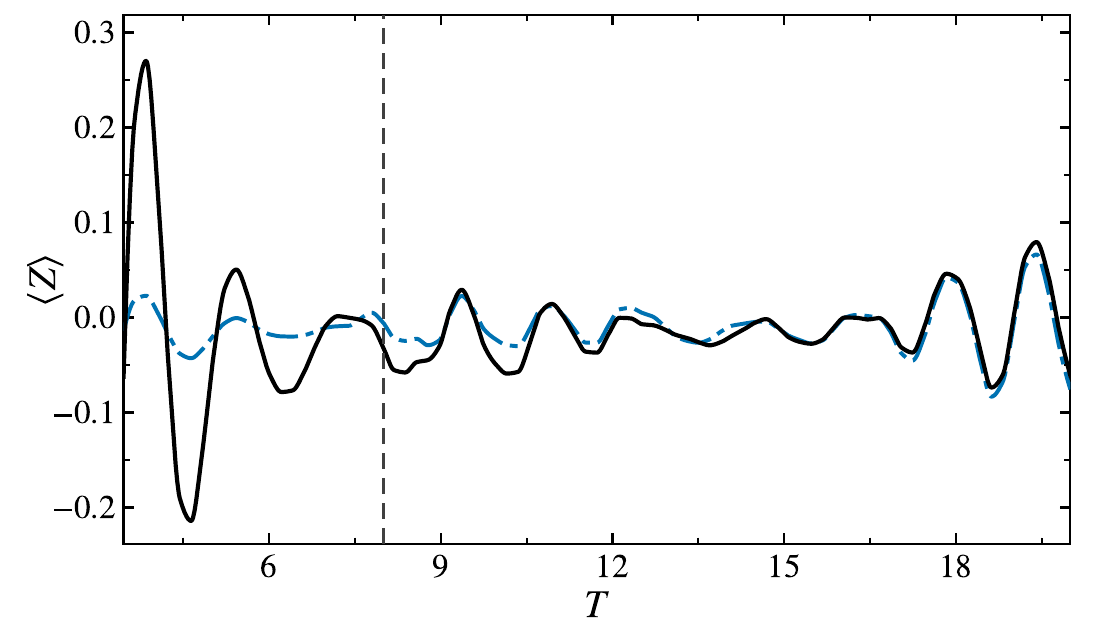}

        \includegraphics[width=\linewidth]{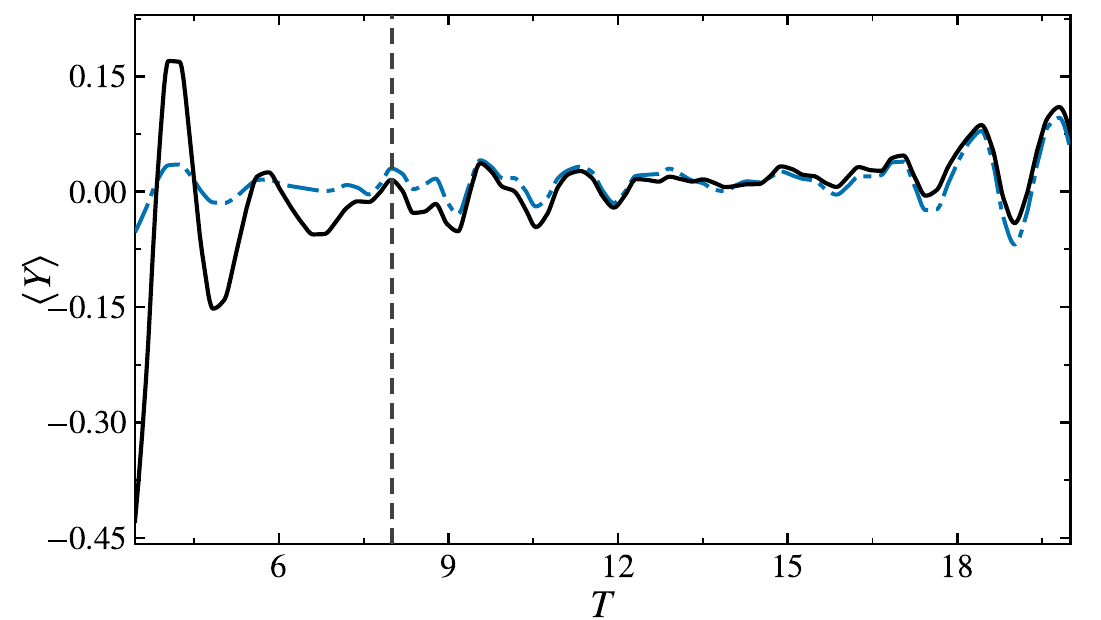}
        \panelabelnumqubit{$N_{q} = 8$}
        \includegraphics[width=\linewidth]{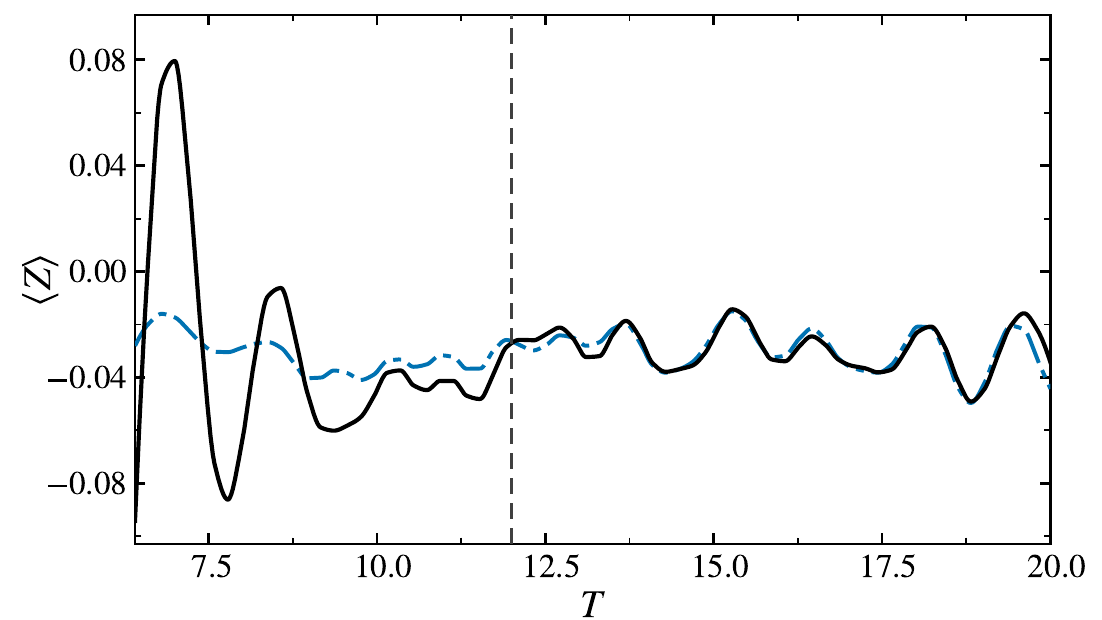}

        \includegraphics[width=\linewidth]{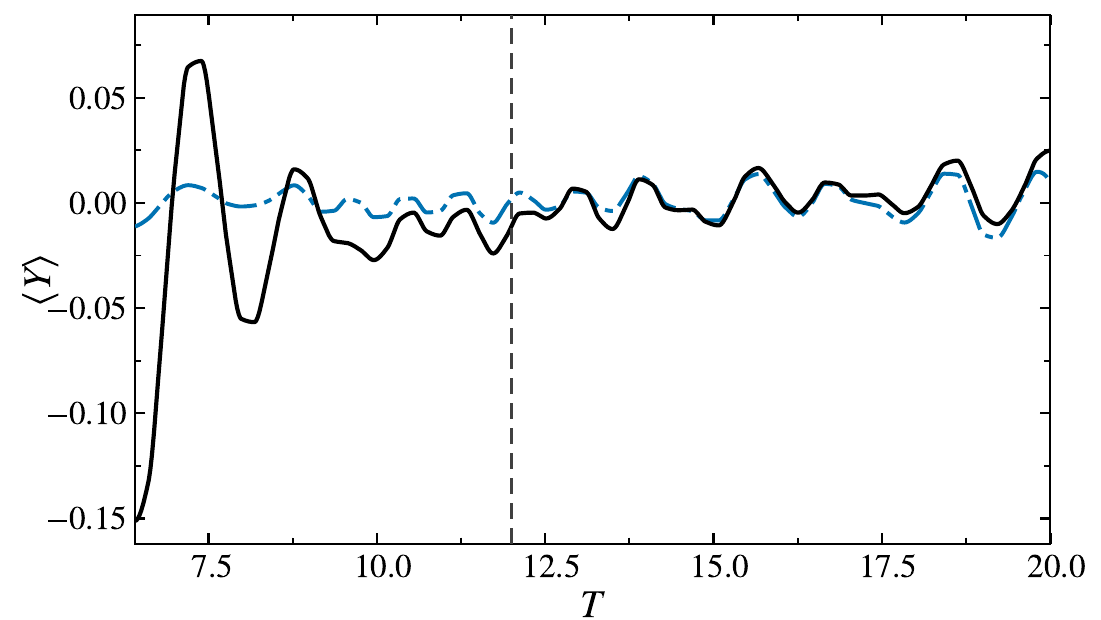}
        \panelabelnumqubit{$N_{q} = 10$}
    \end{minipage}
    \caption{The dynamics of the average magnetizations $\langle Z\rangle$ and $\langle Y\rangle$ in the MFIM model as a function of dimensionless time in the strongly paramagnetic regime $J = h/4$ for 8- and 10-qubit number in fixed-depth Trotter circuits. The left column shows the integrable case $h_{z} = 0$, whereas the right column explores the non-integrable regime, which does not admit a quadratic-fermion description, $|h_{z}| \sim J$. The initial states are $|10110110\rangle$ and $|1011011011\rangle$ respectively, the chain is homogeneous with $J_i=J$, $h_i=h$ and $h_{z,i} = h_{z}$. The dashed line shows the ideal Trotterized evolution for $N=100$ layers. The solid line is the affinely reconstructed noisy signal at gate error probability $p=0.001$ obtained via the same Trotter circuit.}
    \label{fig:MFIM_J=0.25h_big_sim}
\end{figure*}

\begin{figure*}[h]
    \centering

    \begin{minipage}[t]{0.48\textwidth}
        \centering

        \includegraphics[width=\linewidth]{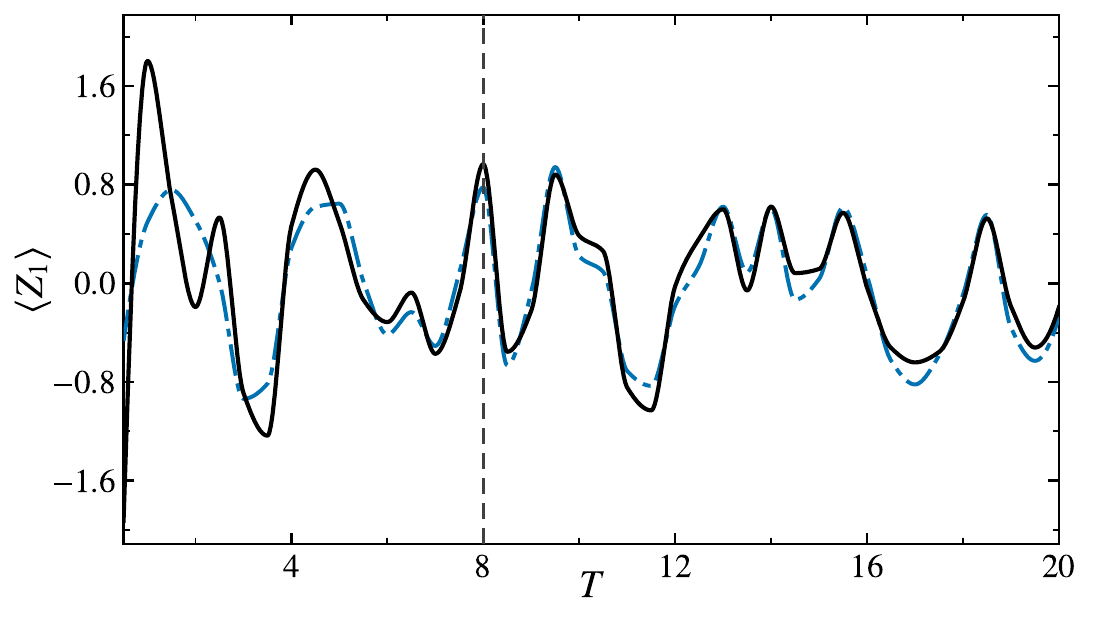}

        \includegraphics[width=\linewidth]{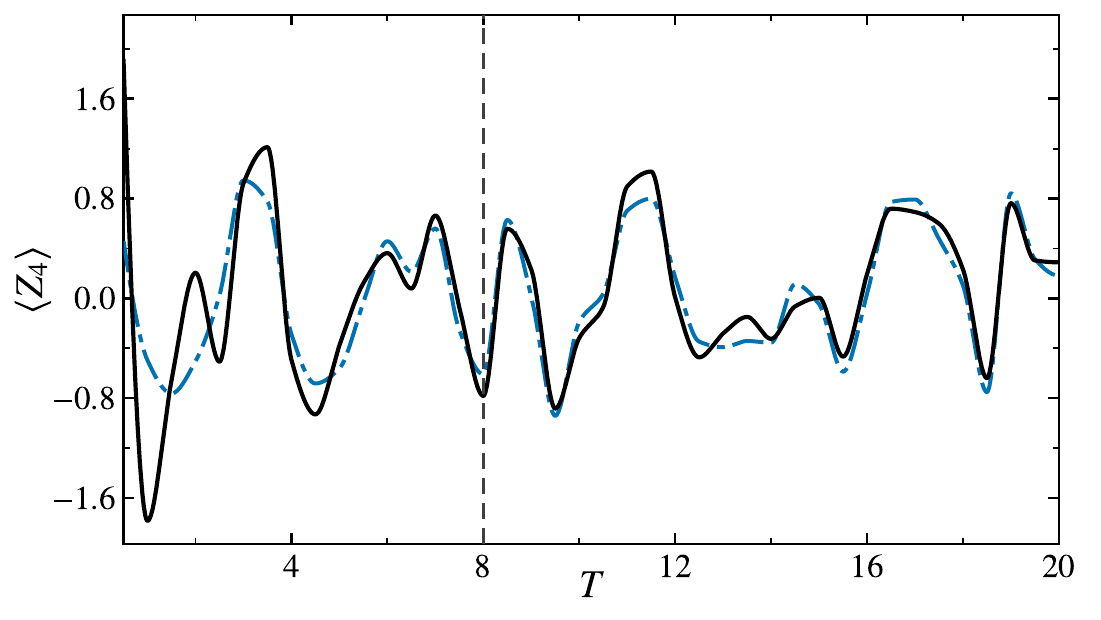}
        \panelabelnumqubit{$N_{q} = 6$}
        \includegraphics[width=\linewidth]{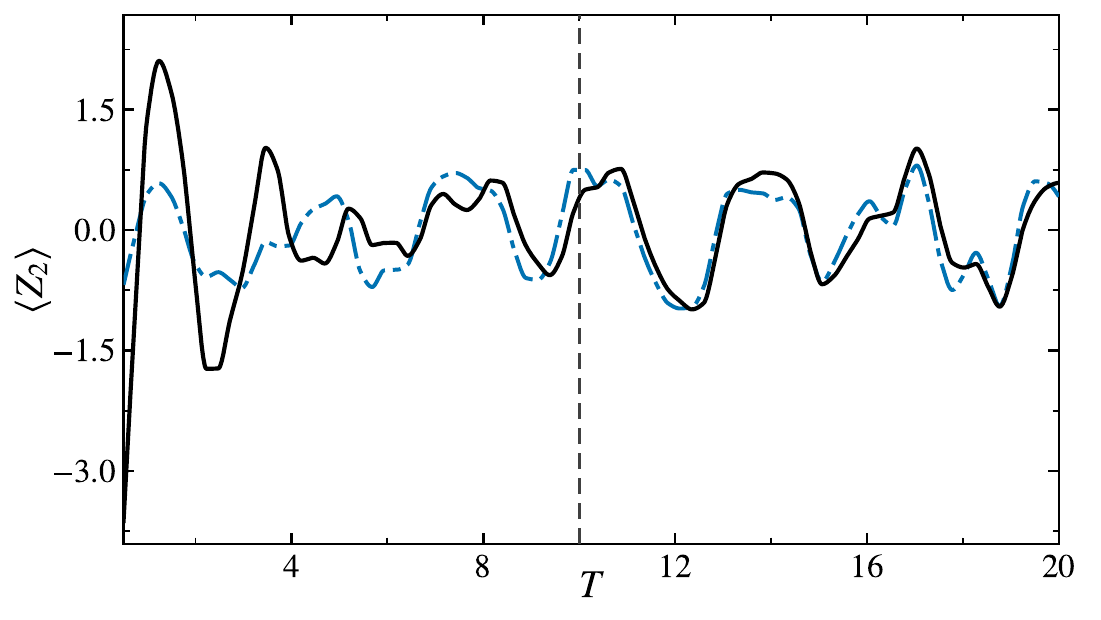}

        \includegraphics[width=\linewidth]{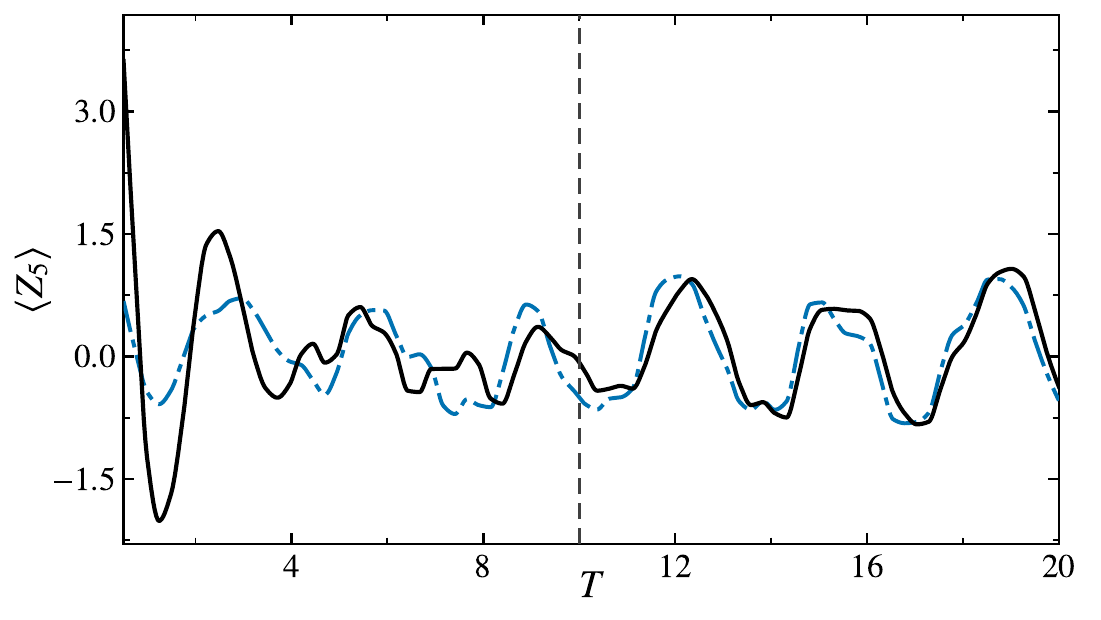}
        \panelabelnumqubit{$N_{q} = 8$}
    \end{minipage}
    \begin{minipage}[t]{0.48\textwidth}
        \centering

        \includegraphics[width=\linewidth]{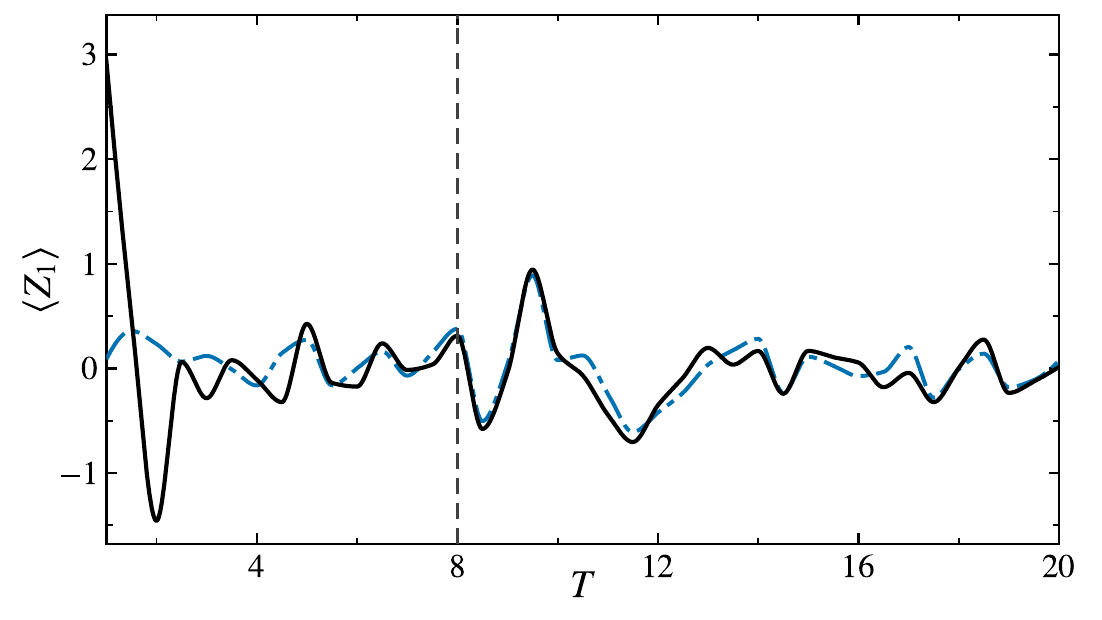}

        \includegraphics[width=\linewidth]{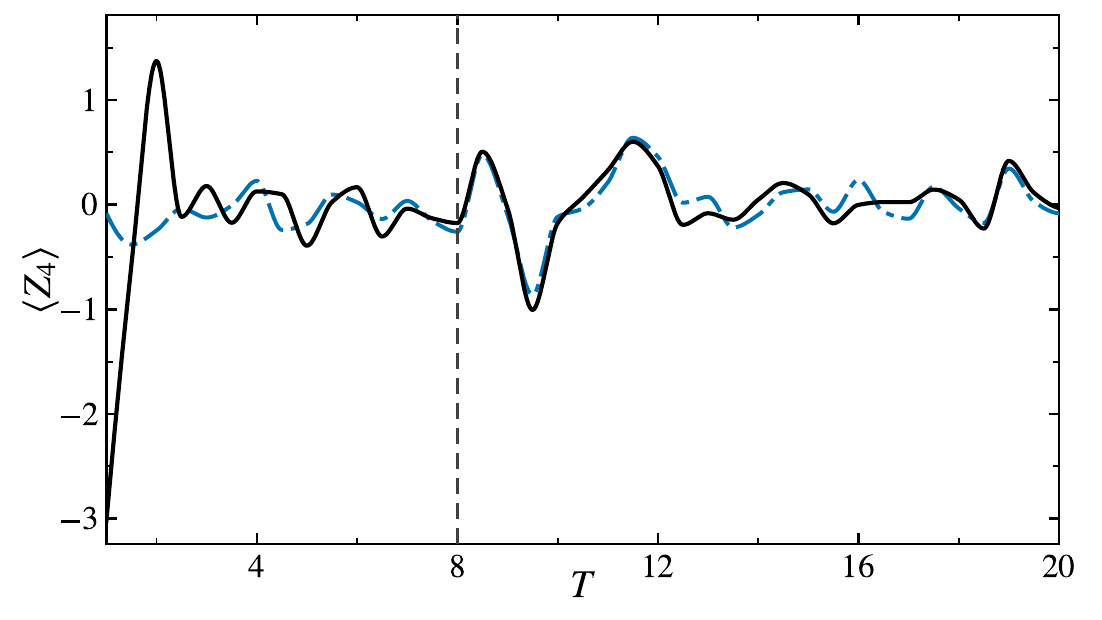}
        \panelabelnumqubit{$N_{q} = 6$}
        \includegraphics[width=\linewidth]{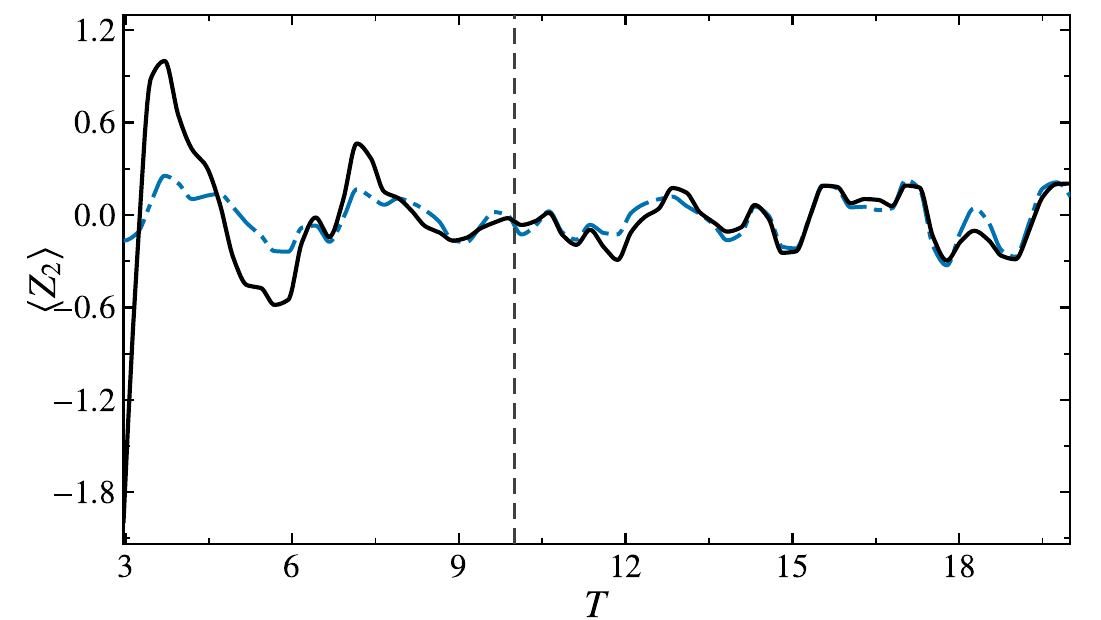}

        \includegraphics[width=\linewidth]{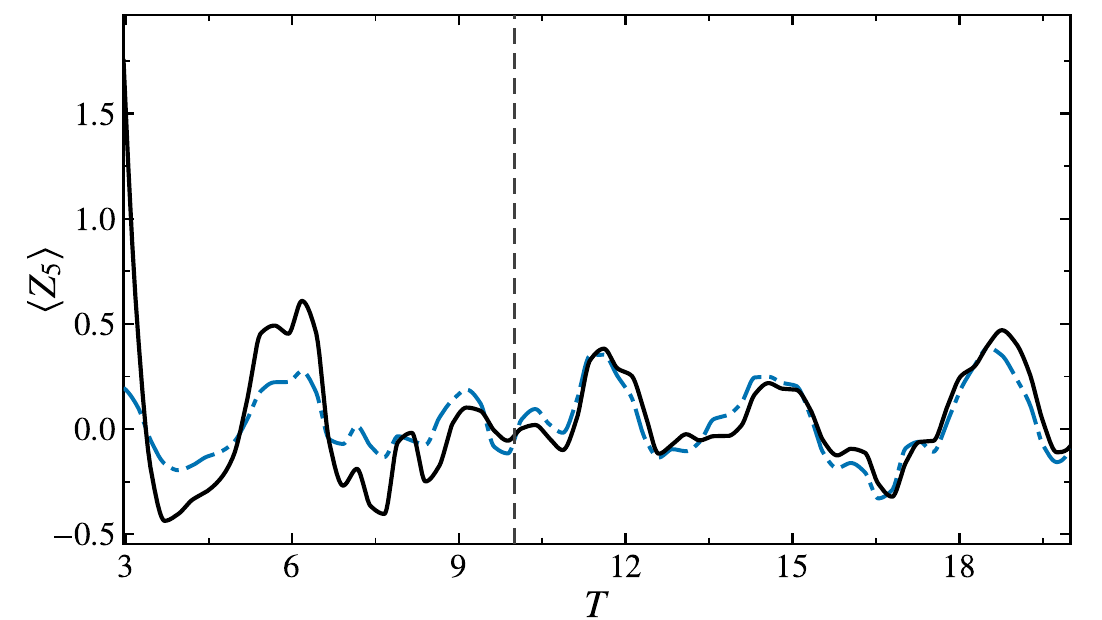}
        \panelabelnumqubit{$N_{q} = 8$}
    \end{minipage}
    \caption{The dynamics of the average on-site magnetization $\langle Z_{j}\rangle$ in the XX transport model as a function of dimensionless time for 6- and 8-qubit number in fixed-depth Trotter circuits. The left column shows the integrable case $V_{I} = 0$, whereas the second column explores the non-integrable regime which is not described in terms of quadratic fermions, $|V_{I}| \sim J/2$. The initial states are the multi-excitation domain-walls $|111000\rangle$ and $|11110000\rangle$ respectively. For each chain, the two monitored sites lie in different initial domains. The dashed line shows the ideal Trotterized evolution for $N=100$ layers. The solid line is the affinely reconstructed noisy signal at gate error probability $p=0.001$ obtained via the same Trotter circuit.}
    \label{fig:XX+YY_big_sim}
\end{figure*}
\end{document}